\documentclass[preprint,tightenlines,aps,floats,nofootinbib,amssymb,groupedaddress]{revtex4-1}
\usepackage[sort&compress]{natbib}
\usepackage{epsf,epsfig}
\usepackage{amsmath}
\usepackage{hyperref}
\usepackage{subfigure}
\usepackage{graphicx}
\usepackage{color}
\usepackage{mathrsfs}
\usepackage{float}

\newcommand{\be}{\begin{equation}}
\newcommand{\ee}{\end{equation}}
\newcommand{\ba}{\begin{array}}
\newcommand{\eaq}{\end{array}}
\newcommand{\bea}{\begin{eqnarray}}
\newcommand{\eea}{\end{eqnarray}}

\newcommand{\nn}{\nonumber}

\newcommand{\bi}{\begin{itemize}}
\newcommand{\ei}{\end{itemize}}
\newcommand{\bal}{\begin{aligned}}
\newcommand{\eal}{\end{aligned}}
\newcommand{\Tr}{\operatorname{Tr}}
\newcommand{\Exp}{\operatorname{Exp}}

\usepackage{comment}
\makeatletter

\begin{document}

\title{Soft Wall holographic model for the minimal Composite Higgs}
\author{Dom\`enec Espriu}
\email{espriu@icc.ub.edu}
\author{Alisa Katanaeva}
\email{katanaeva@fqa.ub.edu}
\affiliation{\it Departament de F\'\i sica Qu\`antica i Astrof\'\i sica and \\
Institut de Ci\`encies del Cosmos (ICCUB), Universitat de Barcelona,\\ 
Mart\'i i Franqu\`es 1, 08028 Barcelona, Catalonia, Spain}

\vspace*{3cm}

\thispagestyle{empty}

\begin{abstract}
We reassess employing the holographic technique to the description of 4D minimal composite Higgs model with $SO(5)\to SO(4)$ global symmetry breaking pattern.
The particular 5D bottom-up holographic treatment is inspired by previous work in the context of QCD and it allows to study spin one and spin zero resonances. 
The resulting spectrum consists of the states transforming under the unbroken $SO(4)$ subgroup and those 
with quantum numbers 
in the $SO(5)/ SO(4)$ coset. The spin one states are arranged in linear radial trajectories, and the states 
from the broken subgroup are generally heavier. The spin zero states from the coset space correspond to the 
four massless Goldstone bosons in 4D. One of them takes the role of the Higgs boson. Restrictions derived 
from the experimental constraints (Higgs couplings, $S$ parameter, etc.) are then implemented and we conclude 
that the model is able to accommodate new vector resonances with masses in the range
$2$ TeV to $3$ TeV without encountering phenomenological difficulties. 
The couplings governing the production of these new states in the processes of the SM gauge boson 
scattering are also estimated. The method can be extended to other breaking patterns.

\end{abstract}

\maketitle

\section{Introduction}
Most of the LHC data gathered so far seems to indicate that the minimal version of the Standard Model (SM)  
with a doublet of complex scalar fields is compatible with the experimental results. However, 
many of the possible extensions  involve a strongly interacting sector where perturbation theory
cannot be trusted and non-perturbative methods are needed to make predictions. The extra-dimensional 
holographic framework is a valid option to investigate strongly coupled theories of various types 
and make meaningful comparisons with experiment.
 
The original AdS/CFT correspondence \cite{Maldacena_1999,Gubser1998,Witten_1998} between string theory 
on $AdS_5\times S_5$  and $\mathcal N=4$ super Yang--Mills gauge theory on $\partial AdS_5$ relates very 
particular theories on both sides. Here, we follow the bottom-up approach to holography -- a conjectured 
phenomenological sprout of AdS/CFT that inherits several key concepts of the latter, but retains 
enough flexibility. It is also known as AdS/QCD due to being tried at and proven successful in  
describing several facets of the SM theory of strong interactions. 

In the AdS/QCD models the spacetime is described by a five-dimensional anti-de Sitter (AdS) metric with 
the additional dimension labelled as $z$. The value $z=0$ corresponds to the ultraviolet (UV) brane, where 
the theory is assumed to be described by a conformal field theory (CFT) as befits QCD at short distances. 
In the infrared (IR) the conformality of the metric must be broken to reproduce the confining property of QCD.
This could be done either via introducing an IR brane at some finite distance in the $z$-direction, 
or making a smooth cut-off  instead.  The former is known as
the hard wall (HW) proposal \cite{HW_2005, daRold_2005}, and the latter is called 
the soft wall (SW) model~\cite{SW_2006} in contrast. The SW framework is of particular phenomenological 
interest  as it results in strongly-coupled resonances lying on linear Regge trajectories.

A viable possibility for an extended electroweak symmetry breaking sector (EWSBS) is the misaligned 
composite Higgs (CH) models~\cite{kaplan84-1, *kaplan84-2, *kaplan84-3, *kaplan84-4, *kaplan85}. Characteristic 
to these models is the breaking of the global symmetry group $\mathcal G$ to a subgroup $\mathcal H$ due to 
some non-perturbative mechanism (like condensation of the fundamental hyper-fermions constructing the 
Higgs and new resonances) at the scale $\Lambda_\text{CH} \simeq 4\pi f_{CH}$. The lightness of the
Higgs is guaranteed by the identification to the  Nambu--Goldstone bosons emerging after the symmetry 
breaking.  The coset space should have capacity for at least four degrees of freedom of the Higgs doublet.

The subgroup $\mathcal H$ should necessarily contain $SU(2)\times U(1)$. However, the SM gauge group 
itself lies in $\mathcal H^\prime$ that is rotated with respect to $\mathcal H$ by a certain angle $\theta$ 
around one of the broken directions. Vacuum misalignment, generated by non-zero $\theta$,  is the mechanism 
responsible for the electroweak (EW)  breaking. Furthermore, the misalignment angle $\theta$ sets the hierarchy 
between $\Lambda_\text{CH}$ and the weak scale $4\pi v$. It is common to assume $v=f_{CH}\sin\theta$. 
One would expect $\sin\theta$ to be small but not too much, because a large scale separation  may 
lead to a relevant  amount of fine-tuning in order to keep light the states that should remain in the 
low energy part of the spectrum.
Moreover, in order to naturally satisfy the constraint on the oblique parameter $T$, $\mathcal H$ should accommodate 
the group of custodial symmetry. 

The Minimal Composite Higgs Model (MCHM) of Ref.~\cite{ACP_2005} provides 
the most economical way to incarnate these demands. It features the groups $\mathcal G=SO(5)$ and 
$\mathcal H=SO(4)\simeq SU(2) \times SU(2)$. Unfortunately, not much is known about the dynamics and the 
spectrum of this theory. The global symmetry $SO(5)$ cannot be realized with 
fermions at the microscopic level. Yet it is often implicitly assumed that a lot of qualitative features 
in CH phenomenology are similar to the ones of QCD. 

There exists substantial bibliography on the application of the holographic methods in CH scenarios. 
One way is to construct a Randall--Sundrum model on a slice of AdS $[z_{UV}, z_{IR}]$. This way minimal 
composite Higgs scenario was realized first in Ref.~\cite{ACP_2005} 
(and followed in Refs.~\cite{AC_2005}, \cite{Medina:2007}, {\it etc.}).  The first example of the technique was proposed
for the simplest case of the $SU(3)\rightarrow SU(2)$ breaking pattern in Ref.~\cite{Contino_Nomura_2003}. 
Other authors used flat 5D models with the $z$ dimension being an orbifold $S^1/Z_2$, {\it i.e.} restricted to  
a finite interval as well (see Refs.~\cite{Panico:2005, Serone2010,Panico:2010}). 

The models inspired by Ref.~\cite{ACP_2005} have the following characteristics. The gauge symmetry of 
the SM is generalized to that of $SO(5)$ and extended into the $5D$ bulk, where the two branes are introduced, 
similar to the HW option in AdS/QCD. The choice of the boundary conditions to be imposed on the $5D$ fields 
on these branes determines the symmetry breaking pattern. 
The Higgs is fully associated with the fifth component of the gauge field in the direction of the broken 
gauge symmetry (an idea first realized in Ref.~\cite{Hosotani1983}).
An effective Higgs  potential  is absent at the tree-level, and its Coleman--Weinberg  generation 
by the quantum loop corrections (dominated by the top quark contribution) breaks the EW  symmetry. 
Emphasis is made on a way one embeds SM quarks into $5D$ model and their impact on the said potential; 
EW observables ($S,\ T,\ Z\rightarrow b\overline{b}$) are also estimated \cite{ACP_2005, AC_2005}. 

CH studies have  not been much elaborated in the SW framework after the initial proposal of Ref.~\cite{Falkowski2008}. 
Motivated by the much better description of QCD phenomenology that SW models provide, we would like to revisit CH
models and provide an in-depth analysis of several relevant observables. We would like to put accent on 
the realization of the global symmetry breaking pattern and the description of spin zero fields, the fulfillment
of the expected current algebra properties, such as Weinberg sum rules, and the OPE. In the present description
the $SO(5) \rightarrow SO(4)$ breaking takes part in the scalar sector of the bulk Lagrangian, similarly
to generalized sigma models used for QCD at long distances~\cite{gasiorowicz, *Fariborz,*Rischke2010}.
The Goldstone bosons are introduced explicitly, but also appear due to the gauge choice in the fifth 
component of the broken gauge field -- that is reminiscent to what was proposed in Ref.~\cite{Falkowski2008}.
However, quite differently from these models, the dynamics responsible for the $SO(5)\to SO(4)$ breaking 
is entirely ``decoupled'' from the SM gauge fields. In our approach, no $SO(5)$ bulk gauge symmetry is  
assumed for the EW sector and only strongly interacting composite states propagate in the bulk. 
The gauge bosons are treated in fact as external sources that do not  participate in the strong dynamics 
(except eventually through mixing of fields with identical quantum numbers) and, hence,  are entirely $z$-independent.
We believe these premises to be well justified after what has been learned from holographic QCD over the last years. The accumulated knowledge vindicates by itself taking another look at CH models.   To specify, our treatment is substantiated by the bottom-up holographic realizations of QCD given in Refs.~\cite{HW_2005, daRold_2005, SW_2006, Hirn2005,Hirn2006, DAmbrosio2015,Espriu_2020}, but several aspects of the $5D$ dynamics are quite distinct for the sake of accommodating the CH physics. 

As said, we concentrate on the dynamics of the strongly interacting EWSBS and its interaction 
with the EW sector, and no new insight into the naturalness problem or the origin of the hierarchy
is provided. We also adopt the point of view that the Higgs potential, being of perturbative origin, is not the primary benefactor of the 
holographic analysis. For that reason we do not introduce SM fermion fields, which in CH scenarios are 
essential to provide the values of $\sin \theta$, Higgs mass and Higgs self-couplings among other 
things~\cite{Bellazzini2014,Panico2016}.

\section{Holographic Composite Higgs framework}
\subsection{Misalignment and operators of the strongly interacting sector}
We will consider a theory where in addition to the SM $\mathcal L_{SM}$ there is a new strongly interacting sector  
$\mathcal {L}_{str. int.}$, presumed to be conformal in the UV. 
A global symmetry of this sector is spontaneously broken following the pattern  $\mathcal G \rightarrow \mathcal H$. 
There are Goldstone bosons in the coset space $\mathcal G / \mathcal H$, and some of them have the quantum numbers 
of the Higgs doublet.
As the $SU(2)_L\times U(1)$  global group  is necessarily included in $\mathcal H$ we can couple the EW sector of the 
SM to the composite sector
\be\label{lagr1}
\mathcal L = \mathcal {\widetilde L}_{str. int.} + \mathcal L_{SM} + \widetilde J_L^{\alpha\ \mu}  W_\mu^{\alpha}
+ \widetilde J^{Y\ \mu}  B_\mu.
\ee
There only appear the conserved currents of the
strongly interacting sector $J_L^{\alpha\ \mu}$ and $J^{Y\ \mu}$ that contain the generators of the EW group.  
Moreover, we have to denote the misalignment between the $\mathcal H$ subgroup of the new sector and the actual 
$\mathcal H'$ containing the $W_\mu^{\alpha}$ and $B_\mu$ EW gauge bosons. 
In Eqn. (\ref{lagr1}) everything related to the new composite sector is marked with tildes.

Let us specify to the case of MCHM, where the global symmetry breaking pattern 
is $SO(5) \rightarrow SO(4)$ and there are exactly four Goldstones.
We denote by $T^A, \ A=1,...,10$ the generators of $SO(5)$, 
represented  by $5\times5$ matrices, which are traceless $\Tr T^A=0$ and normalized as $\Tr (T^AT^B)=\delta^{AB}$. 
They separate naturally into two groups:
\bi
\item The unbroken generators, in the case of MCHM those of $SO(4)\backsimeq SU(2)_L\times SU(2)_R$, we will cal
$T^a, \ a=1,...,6$. They are specified as
\be \label{genso4}
    T^\alpha_L=\begin{pmatrix}
    t^\alpha_L & 0\\
    0 & 0\\
   \end{pmatrix},\ T^\alpha_R=\begin{pmatrix}
    t^\alpha_R & 0\\
    0 & 0\\
   \end{pmatrix}, \ \alpha=1,2,3,
\ee
where $t^\alpha_{L}$, $t^\alpha_{R}$ are $4\times4$ matrices given by 
$(t^\alpha_{L/R})_{jk}=-\frac i2 (\varepsilon_{\alpha\beta\gamma}\delta^\beta_j\delta^\gamma_k
\pm (\delta_j^\alpha\delta_k^4-\delta_k^\alpha\delta_j^4))$, $j,k=1,...,4$.
\item The broken generators, corresponding to the coset $SO(5)/SO(4)$, are labeled as 
$\widehat{T}^i,\  i=1,2,3,4$ and defined as follows 
\be\widehat{T}^i_{IJ}=-\frac i{\sqrt 2}(\delta^i_I\delta^5_J-\delta^i_J\delta^5_I),\ \ I,J=1,...,5. \ee
\ei

A quantity parametrizing  the vacuum misalignment and responsible for the EW symmetry breaking is the rotation 
angle $\theta$ that relates the linearly-realized global group $\mathcal H=SO(4)$
and the gauged  group $\mathcal H'=SO(4)'$. 
It is natural to assign the value $\theta=0$ to the SM, hence we denote the generators of $SO(5) \rightarrow SO(4)'$ 
as $\{T^a(0),\ \widehat{T}^i(0)\}$
and those of $SO(5) \rightarrow SO(4)$ as $\{T^a(\theta),\ \widehat{T}^i(\theta)\}$.
We choose a preferred direction for the misalignment and the following connection between the generators holds
\be \label{rotated_gen}
T^\alpha(\theta)=r(\theta)T^\alpha(0)r^{-1}(\theta), \ \text{with} \ r(\theta)=\begin{pmatrix}
	1_{3\times3} & 0 & 0\\
	0 &\cos\theta&\sin\theta\\
	0 & -\sin\theta&\cos\theta\\
\end{pmatrix}.
\ee

Compositness implies that some fundamental degrees of freedom are bound together by the new ``color'' force 
(hyper-color is usually used in the CH framework).
MCHM does not admit complex Dirac fermions as fundamental fields at the microscopic level due to the nature
of the global ``flavor'' symmetry group. The anomaly-free UV complete fundamental fermion theory should have 
$\mathcal G$  equivalent to $SU(n_1)\times \ldots \times SU(n_p)\times U(1)^{p-1}$, where $n_i$ is the number of 
fermions in a given irreducible representations and $p$ counts the number of different irreps \cite{Ferretti2014}. 
The simplest UV-completable theory will be the next-to-minimal CH with $SO(6)\rightarrow SO(5)$, featuring five 
Goldstone bosons (other next-to-minimal patterns are mentioned, for instance, in Ref.~\cite{Cacciapaglia2014}). 
Nevertheless, we choose to work with MCHM because of its simplicity that serves to illustrate the general procedure.

If one chooses to avoid the particularities of the microscopic structure of the new composite states (that 
seems advisable on the grounds of being as general as possible), it is 
impossible to treat the holographic MCHM completely in the AdS/QCD fashion of Ref.~\cite{HW_2005, SW_2006}.
To some extent, due to affecting directly the operator scaling dimension $\Delta$, the microscopic substructure sets 
the prescriptions for the bulk masses and UV boundary conditions, 
 which in their turn influence all other holographic derivations.
In our holographic model describing the minimal CH we only use a single entry from the list of field-operator 
correspondences \cite{Aharony2000}
\be\label{vector_field_dual}
A_\mu^A(x,z=\varepsilon)=1\cdot \phi_\mu^A(x) \ \leftrightarrow \ \mathcal O_\mu^A(x)\ \text{with}\ \Delta=3,
\ee
where $\mathcal  O_\mu^A(x)$ are the unspecified conserved currents of the fundamental theory containing $SO(5)$ 
generators $T^A$, and $A_\mu^A(x,z)$ are dual $5D$ fields restricted to provide the sources $\phi_\mu^A(x)$ 
for the corresponding operators on the UV brane  ($\varepsilon$ is an UV regulator). We take $\Delta=3$ (and zero 
bulk mass of the vector fields) as a universal feature for the conserved vector currents, because it should be so 
both in the case of fermionic ($\overline \Psi\gamma_{\mu}T^A \Psi$) and bosonic ($\partial_\mu s^\top T^A s$) 
fundamental degrees of freedom. The introduction of the scalar operator is indispensable in order to generate 
the breaking towards $SO(4)$. However, following a line similar to the vector case would  mean inferring too much 
on the nature of the fundamental theory. Hence, we intend to construct the model so that this part of duality is 
realized in an alternative way. 
	
The operators $\mathcal  O_\mu^A(x)$ define the currents of Eqn.~(\ref{lagr1}):
\bi
\item for $A=\alpha$ (left): $\frac{g}{\sqrt2 }\mathcal O^{\alpha}_{L\mu}(x)=g_V J^{\alpha}_{L\mu}$;
\item for hypercharge realized  as $Y=T^3_{R}$: $\frac{g'}{\sqrt2}\mathcal O^{3}_{R\mu}(x)=g_V J^Y_{\mu}$.
\ei
The coupling coefficients are not fully established because the operators are taken with an abstract normalization $g_V$ 
that will be determined to provide agreement with the common MCHM notations.
Introduction of $g_V$ is also substantiated by the discussion of Ref.~\cite{Espriu_2020}, where it is argued that 
a degree of arbitrariness in the field-operator holographic correspondence is a necessary piece of 
AdS/QCD constructions.

\subsection{5D model Lagrangian}
\label{sec-model}
In this subsection we put forward the details of the holographic 5D model realizing the 4D MCHM concept.
We settle upon the idea that there are two composite operators, a vector and a scalar one, that define the theory, 
and hence we have spin one and spin zero fields on the 5D side. These fields live in the 5D AdS bulk with a metric 
given by
\be
g_{MN}dx^M dx^N=\frac{R^2}{z^2}(\eta_{\mu\nu}dx^\mu dx^\nu-d^2z),\quad \eta_{\mu\nu}=\text{diag}(1,-1,-1,-1).
\ee
The dynamics is governed by the following $SO(5)$ gauge invariant action
\begin{align}\label{5Daction}
S_{5D}=&\frac{1}{4g_5^2}\int d^5x\sqrt{-g}e^{-\Phi(z)}\Tr F_{MN}F_{KL}g^{MK}g^{LN}\\
&+\frac1{k_s}\int d^5x\sqrt{-g}e^{-\Phi(z)}\bigg[ \Tr g^{MN}(D_MH)^\top(D_NH)
-M^2_H \Tr HH^\top \bigg].\nn
\end{align}
This 5D effective action includes matrix-valued scalar and vector fields and, as mentioned, is inspired 
by generalized sigma models used in the context of strong interactions. A similar starting action 
was used in the AdS/QCD study of Ref.~\cite{Espriu_2020}. The dimensionality of the normalization 
constants $g_5^2$ and $k_s$ is set to compensate that of the additional dimension: $[g_5^2]=[k_s]=E^{-1}$. 
To have the gravitational background of a smoothly capped off AdS spacetime we  introduce a SW 
dilaton function $\Phi(z)=\kappa^2z^2$ in the common  inverse exponent  factor.

The scalar degrees of freedom are collected in the matrix-valued field $H$.
Let us denote the group transformations $g\in SO(5)$ and $h\in SO(4)$.
The matrix of the Goldstone fields $\xi$ transforms under $SO(5)$ as: $\xi\rightarrow \xi'=g\xi h^\top$. The other scalar 
degrees of freedom with the quantum numbers of SO(4) are collected in the matrix $\Sigma$ transforming as 
$\Sigma\rightarrow\Sigma'=h\Sigma h^\top$. The breaking from $SO(5)$ to $SO(4)$ also appears there and is parametrized by
a function $f(z)$. From these components we can construct a proper combination leading to $H\rightarrow H'=g H g^\top$
\be
H=\xi\Sigma\xi^{\top},\ \Sigma=\begin{pmatrix}
	0_{4\times4} & 0\\
	0 & f(z)\\
\end{pmatrix}+\sigma^a (x,z)T^a,
\ \xi=\exp\left(\frac{i\pi^i(x,z)\widehat{T}^i}{\chi_\pi}\right),
\ee
where $[\chi_\pi]=[f(z)]=E^1$. The minutiae of the scalar fields, introduced in $\Sigma$ as $\sigma^a$, 
will be further omitted in this study.
It follows then that in this representation: $H=H^\top$,  the $\Tr HH^\top$ quadratic piece of Eqn.~(\ref{5Daction}) brings no field interactions 
and the value of $M^2_H$ is of no consequence.

Holography prescribes that every global symmetry of the 4D model comes as a gauge symmetry of its 5D dual.
Thus, to make the Lagrangian invariant under the gauge transformation
$A_M \rightarrow A'_M=gA_M g^{-1}+ig\partial_M g^{-1}$
the covariant derivative is introduced in the 5D action~(\ref{5Daction}), defined as
\be
D_MH=\partial_M H+[A_M, H], \quad D_MH\rightarrow g D_MH g^{-1}.
\ee
The field strength tensor that produces the vector field kinetic term in Eqn.~(\ref{5Daction}) is  
\be F_{MN}=\partial_MA_N-\partial_NA_M+[A_M, A_N].
\ee
Generally, we take $A_M=-iA_M^AT^A$, where the upper index runs through both broken and unbroken indices 
$A_M^aT^a+A_M^i\widehat{T}^i$. These 5D vector fields are unrelated to the $W_\mu^\alpha$ or $B_\mu$ gauge 
bosons of the EW interactions, but for their eventual mixing.

The $A_\mu^A$  fields are connected by duality to the  $\mathcal O_\mu^A$ vector composite operators with the same generators and have the boundary condition~(\ref{vector_field_dual}).
For the fifth component of the vector field we assume that
\be \label{Az_boundary}
A_z^A(x,\varepsilon)=0,\ee
because there is no $4D$ source for it to couple to. The common holographic gauge $A_z^A\equiv 0$ fulfills 
this condition trivially, but this is not the only possibility. On the other hand, the dual counterpart 
of $H$ and the value of $M^2_H$ remain unspecified. The near-boundary behavior of the Goldstone fields 
$\pi^i(x,z)$ will be eventually determined in Section \ref{sect:EOM_broken} from considerations of another type. The treatment of the Goldstone 
is an essential aspect  of the model because they correspond to the four components of the Higgs doublet.

\subsection{Extraction of $4D$-relevant physics}
The basic principle of AdS/CFT correspondence states that the partition function of the 4D theory and the 
on-shell action of its 5D holographic dual coincide in the following sense \cite{Witten_1998, Gubser1998}:
\be\label{cor_corresp}
Z_{4D}[\phi]=\Exp i S_{5D}^{on-shell}|_{\phi(x,z)\rightarrow\phi(x,z=\varepsilon)}.
\ee
Essentially, all bulk fields $\phi(x,z)$ are set to their boundary values $\phi(x,z=\varepsilon)$, which 
could be identified with the sources $\phi(x)$ as in the case of Eqn.~(\ref{vector_field_dual}).

The dynamics of holographic fields is governed by a set of second order equations of motion (EOMs). Thus, 
a 5D field can be attributed with two solutions.
According to the usual AdS/CFT dogma, the leading mode at small $z$ corresponds to the bulk-to-boundary propagator. 
It connects a source at the boundary and a value of a field in the bulk and should exhibit enough decreasing 
behavior in the IR region to render the right-hand side of Eqn.~(\ref{cor_corresp}) finite. 
The subleading mode represents an infinite series of normalizable solutions, known as the Kaluza-Klein (KK) 
decomposition. There, the 4D and $z$ dependencies are separated: the $z$-independent functions are identified 
with a tower of physical states at the 4D boundary that are further promoted into the bulk with the 
$z$-dependent profiles.

\begin{figure}\center
\includegraphics[scale=0.25]{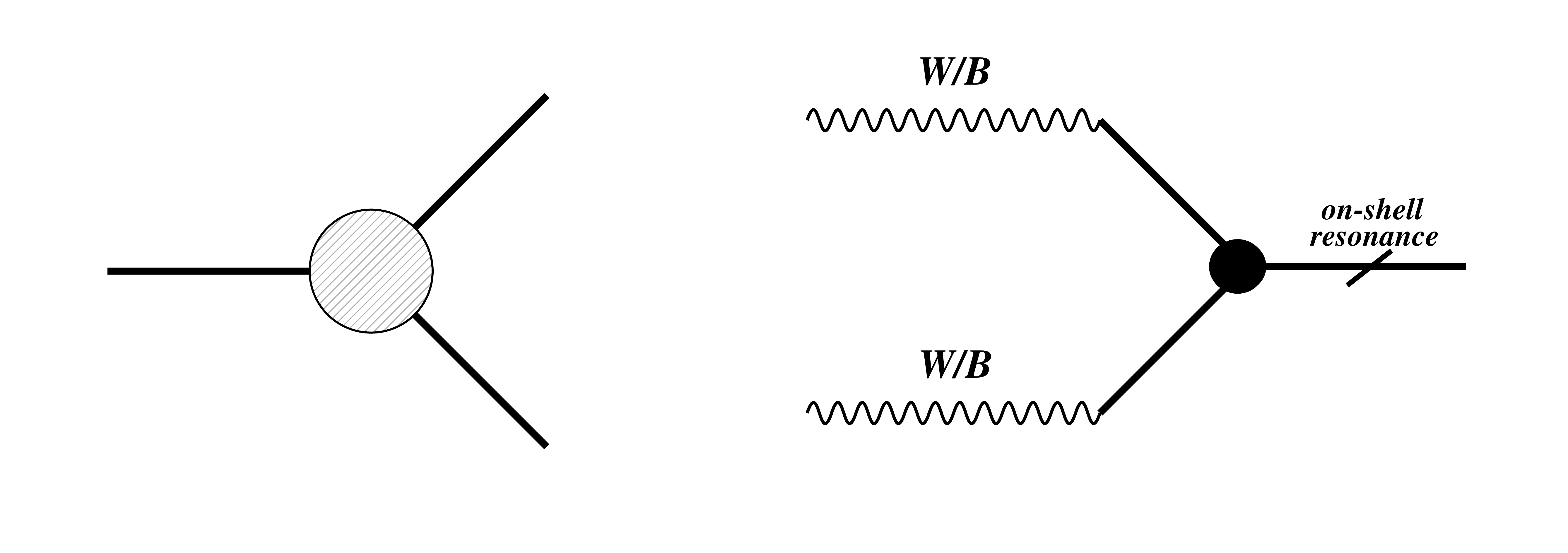}
	\caption{\label{effective_coupl} Diagrams describing (left) three-point correlation function, (right) 
effective triple couplings between two SM gauge bosons and a composite resonance.}
\end{figure}

From consideration of the KK solutions one gets knowledge about the spectra of the composite 4D resonances.
While from Eqn.~(\ref{5Daction}), evaluated on the bulk-to-boundary solutions, one can extract the $n$-point 
correlation functions of the composite operators \cite{Witten_1998, Gubser1998, Freedman1999}. 
The 4D partition function is given by the functional integral over the fundamental fields $\varphi$
contained in the selected operators ({\it e.g.} $\mathcal O^{A}_\mu$) and in the fundamental Lagrangian 
$\mathcal L_{str. int.}$
\bea
Z_{4D}[\phi]&=&\int [\mathcal D\varphi]\Exp i\int d^4x [\mathcal L_{str. int.}(x)+ 
\phi_{\mu}^A(x)\mathcal O^{A\mu} (x)
+\ldots] \nn \\ 
&=&\Exp\sum\limits_q\frac1{q!}\int\prod\limits_{k=1}^qd^4x_k\langle\mathcal O_1(x_1)...\mathcal O_q(x_q)\rangle
i\phi^1(x_1)...i\phi^q(x_q).\label{Zqft}
\eea
From the schematic definition in Eqn.~(\ref{Zqft}) and the correspondence postulate~(\ref{cor_corresp}), 
it is clear that the Green functions can be obtained by the variation of the 5D effective action with respect 
to the sources. Diagrammatically we can represent the correlation functions by the left panel of 
Fig.~\ref{effective_coupl}, where in general the number of legs could be equal to the number $n$ of 
operators in the correlator.
At the same time, couplings involving the composite resonances can be estimated taking the proper term 
in the 5D Lagrangian, inserting the KK modes for the interacting 5D fields and integrating over the 
$z$-dimension. Due to $\ln Z_{4D} = i S^{eff}_{4D}$ a calculation of this kind brings an effective vertex.

Interaction of a given composite state with the SM gauge bosons happens through the mixing of the 
latter with other composite particles. Due to the misalignment the EW bosons couple to a variety of 
resonances, because the rotated currents $\widetilde J^\alpha_\mu$ overlap with different types 
of vectorial currents that are holographically connected to vector composite fields. Besides, 
all radial excitations in a KK tower should generally be included in the internal propagation. 
The procedure in this case is the following: calculate the $n$-point correlation function, build the 
effective 4D Lagrangian via attaching $W_\mu^{\alpha}$ or $B_\mu$ fields as physical external sources, 
and reduce the legs where the composite resonances become physical and put on-shell (substituted with 
their KK modes). This is shown in the right panel of Fig.~\ref{effective_coupl}.

\section{Equations of motion and their solutions}
In this section we study the EOMs of the 5D fields. They are derived from the $5D$ action  at the quadratic level
\begin{align}\notag
S^{(2)}_{5D}=&\int d^5xe^{-\Phi(z)}\left\{-\frac{1}{4g_5^2}\frac RzF^{ A}_{\mu\nu}F^{ A\mu\nu}+\frac{1}{2g_5^2}
\frac{R}{z}(\partial_z A^A_\mu-\partial_\mu A^A_z)(\partial_z A^{A\mu}-\partial^\mu A^A_z) \right.\\ \label{lagr2}
&+\left.\frac{f^2(z)}{k_s}\frac{R^3}{z^3}\left[\left(A^i_\mu-\partial_\mu\frac{\pi^i}{\chi_\pi}\right)^2- 
\left(A^i_z-\partial_z\frac{\pi^i}{\chi_\pi}\right)^2\right]\right\}.
\end{align}
The sum over coincident indices is assumed for $A=\{a,i\}=1,\ldots,10$ in the first line, and just over 
broken indices $i=1,\ldots,4$ in the second. The ansatze functions are $\Phi(z)=\kappa^2z^2$ and $f(z)\sim z$. 
The choice for the symmetry breaking function $f(z)$ is justified by the analyticity of the solution 
in the broken vector sector; the argumentation is similar to that of Ref.~\cite{Espriu_2020}.

\subsection{Equations of motion for the unbroken generators}
In  the unbroken sector with ${a}=1,..,6$
\bea\label{unbr_Amu}
\partial_z \frac{e^{-\Phi(z)}}{z}\partial_z A_\mu^a-\frac{e^{-\Phi(z)}}{z}\Box A_\mu^a-
\partial_z \frac{e^{-\Phi(z)}}{z}\partial_\mu A_z^a=0, \\
\Box A_z^a=\partial^\mu\partial_z A_\mu^a.
\eea
If we act with $\partial^\mu$ on the first equation and substitute $\Box A_z^a$ from the second one, 
we would get the third term equal to the first one. Then, the result is
\be \label{unbroken_cond}
\Box \partial^\mu A_\mu^a=0,
\ee
that implies either $\partial^\mu A_\mu^a=0$ (transversality) or $q^2_{A^\parallel}=0$ (longitudinal mode), where
\be
A_\mu^a=A_\mu^{a\bot}+A_\mu^{a\parallel},\ee
with $A_{\mu}^{a\bot}=\mathcal P_{\mu\nu}A^{a\nu}$, $\mathcal P_{\mu\nu}=\left(\eta_{\mu\nu}-\frac{q_\mu q_\nu}{q^2}\right)$, 
and $A_\mu^{a\parallel}=\frac{q_\mu q_\nu}{q^2}A^{a\nu}$. 

The condition~(\ref{unbroken_cond}) modifies the second equation in the system into
\be \label{unbr_Az}
\Box^2 A_z^a=0.
\ee
While acting with $\Box^2$ on Eqn.~(\ref{unbr_Amu}) and taking into account $q^2_{A^\bot}\neq0$ we get 
the following equation for the transversal mode
\be \label{unbr_tr_EOM1}
\partial_z \frac{e^{-\Phi(z)}}{z}\partial_z A_\mu^{a\bot}-\frac{e^{-\Phi(z)}}{z}\Box A_\mu^{a\bot}=0.
\ee
However, the result for the longitudinal mode with $q^2_{A^\parallel}=0$ turns out trivial, meaning that the 
remaining system for $A_\mu^{a\parallel}$ and $A_z^a$ is underdefined. 
We choose to work in a class of solutions where Eqn.~(\ref{unbr_Az})  is fulfilled  with the gauge
\be
A_z^a(x,z)\equiv0.
\ee
As a result the EOM for the longitudinal mode simplifies to
\be\label{unbr_l_EOM1}
\partial_z A_\mu^{a\parallel}=0.
\ee

The following boundary terms are left in the on-shell action~(\ref{lagr2})
\be
\frac1{2g_5^2} \int d^4x \left. e^{-\Phi(z)} \frac Rz A^{a\mu} (\partial_z A^a_\mu- \partial_\mu A^a_z)\right|_\varepsilon^\infty= -\frac1{2g_5^2}  \int d^4x \left.  \frac Rz A^{a\bot\mu} \partial_z A^{a\bot}_\mu\right|_{z=\varepsilon}.\label{boundary_unbr}
\ee
Only the transversal term remains, giving rise to the two-point function studied in 
Section~\ref{sec-corr-vector}. 

Let us perform a 4D Fourier transform  $A_\mu^a(x,z)=\int d^4q e^{iqx}A_\mu^a(q,z)$ and
let us focus on finding solutions of the EOMs. 
 First, the transverse bulk-to-boundary propagator, 
which we denote $V(q,z)$, is defined by
\be
A_\mu^{a\bot}(q,z)=\phi_{\mu}^{a\bot}(q)\cdot V(q,z),\quad\ V(q,\varepsilon)=1,
\ee
where $\phi_{\mu}^{a\bot}$ should be understood as a projection of the original source 
$\phi_{\mu}^{A\bot}=\mathcal P_{\mu\nu}\phi^{A\nu}$. The analogous longitudinal 
projection will be denoted by $\phi_{\mu}^{A\parallel}$.

From Eqn.~(\ref{unbr_tr_EOM1}), changing to the variable $y=\kappa^2z^2$, we arrive to the following EOM
\be
y V''(q,y)-yV'(q,y)+\frac{q^2}{4\kappa^2}V(q,y)=0
\ee
It is a particular case of the confluent hypergeometric equation (see Appendix~\ref{hyperf} 
for a review of the properties and solutions of this equation), and the dominant mode at small $z$  is
\be\label{vprop}
V(q, z)=\Gamma\left(-\frac{q^2}{4\kappa^2}+1\right)\Psi\left(-\frac{q^2}{4\kappa^2},0;\kappa^2z^2\right).
\ee

The subdominant solution (see Eqn.~(\ref{hyperfsol})) gives us the tower of massive states, identified 
with vector composite resonances at the boundary. 
Normalizable solutions can only be found for  discrete values of the 4D momentum
$q^2=M_V^2(n)$ and we may identify $\left.V(q,z)\right|_{q^2=M_V^2(n)}=V_n(z)$.
The KK decomposition is set as follows
\be A^{a\bot}_\mu(q,z)=\sum\limits_{n=0}^\infty V_n(z)A_{\mu (n)}^{a\bot}(q).\ee
The $z$ profile and the spectrum can be expressed using the 
discrete parameter $n=0,1,2,...$
\be\label{vmodes}
V_n (z)=\kappa^2z^2 \sqrt{\frac{g_5^2}R}\sqrt\frac{2}{n+1} L_n^1(\kappa^2z^2),\quad \
M_V^2(n)=4\kappa^2(n+1),
\ee
where $L_n^m(x)$ are the generalized Laguerre polynomials. The profiles $V_n(z)$ are subject
to the Dirichlet boundary condition and are normalized to fulfill the 
orthogonality relation
\be \frac R{g_5^2} \int\limits_0^\infty dz e^{-\kappa^2z^2}z^{-1}V_n(z)V_k(z)=\delta_{nk}.\label{unbr_orthog}\ee

For the longitudinal mode, $A_\mu^{a\parallel}(q,z)$, the bulk-to-boundary solution is similarly 
defined. Its EOM~(\ref{unbr_l_EOM1}), however, admits only trivial continuation into the bulk
\be
A_\mu^{a\parallel}(q,z)=\phi_\mu^{a\parallel}(q)\cdot V^{\parallel}(q,z),\quad V^\parallel(q, z)=1.
\ee

The previous results are well known. Let us now see the equivalent derivation in the broken sector.

\subsection{Equations of motion for the broken generators}\label{sect:EOM_broken}
The EOMs for {the broken sector} with ${i}=1,..,4$ are more complicated due to the appearance of mixing with $\pi^i$
\begin{align}
&\partial_z \frac{e^{-\Phi(z)}}{z}\left(\partial_z A^i_\mu-\partial_\mu A^i_z\right)-\frac{e^{-\Phi(z)}}{z}\Box A^i_\mu-\frac{2g_5^2 f^2(z)R^2}{k_s}\frac{e^{-\Phi(z)}}{z^3}\left(A^i_\mu-\frac{\partial_\mu \pi^i}{\chi_\pi}\right)=0 \label{EOM1}\\
&\frac{e^{-\Phi(z)}}{z}\left(\partial^\mu\partial_z A^i_\mu-\Box A^i_z\right)-\frac{2g_5^2 f^2(z)R^2}{k_s}\frac{e^{-\Phi(z)}}{z^3}\left(A^i_z-\partial_z\frac{\pi^i}{\chi_\pi}\right)=0 \label{EOM2}\\
&\partial_z \frac{f^2(z)R^2e^{-\Phi(z)}}{z^3}\left(A^i_z-\partial_z\frac{\pi^i}{\chi_\pi}\right)
-\frac{f^2(z)R^2e^{-\Phi(z)}}{z^3}\left(\partial^\mu A^i_\mu - \Box \frac{\pi^i}{\chi_\pi}\right)=0 \label{EOM3}
\end{align}
Combining $\partial^\mu\times$~(\ref{EOM1}) with other two equations we arrive again at the condition 
\be
\Box \partial^\mu A^i_\mu=0, \label{trans_cond}
\ee
with the  same options $\partial^\mu A_\mu^i=0$ and $q^2_{A^\parallel}=0$ as in the unbroken case. 
The condition on $A^i_z$ is different though
\be
\partial_z \frac{e^{-\Phi(z)}}{z}\Box^2 A^i_z-\frac{2g_5^2f^2(z)R^2e^{-\Phi(z)}}{k_sz^3} 
\Box^2 \frac{\pi^i}{\chi_\pi}=0 \label{broken_cond}.
\ee

The system of equations obeyed by $A_\mu^{i\parallel}$, $A_z^i$ and $\pi^i$ is insufficient to determine them and we 
can only solve the problem with the help of an appropriate gauge condition. There are various possibilities, 
but we find the option explained below  most useful for the physics we aspire to describe. We impose
\be \label{Az_of_xi}
A^i_z(x,z)=\xi \partial_z \frac{\pi^i(x,z)}{\chi},
\ee
where the parameter $\xi$ is arbitrary.
 
The fact that $\pi^i(x,z)$ appears both in the scalar part of the model Lagrangian and in this gauge 
condition makes  it distinct from other $5D$ fields in the model. To analyze the Goldstone solution we assume 
that the corresponding EOM defines the $z$-profile $\pi(x,z)$ that couples to the physical 
mode $\pi^i(x)$ on the boundary. The Neumann boundary condition, $\partial_z \pi(x,z)|_{z=\varepsilon}=0$, 
is imposed due to Eqn.~(\ref{Az_boundary}). 

Now both parts of Eqn.~(\ref{broken_cond}) have the same $x$-dependence, and $\Box^2$ can be taken 
out of the bracket. It results in the following equation on $\pi(x,z)$
\be\label{pionEOM}
\partial_z \frac{e^{-\Phi(z)}}{z}\partial_z \pi(x,z)-\frac{2g_5^2f^2(z)R^2}{\xi k_s}\frac{e^{-\Phi(z)}}{z^3}\pi(x,z)=0.
\ee
At the same time it allows to get rid of $A^i_z$ and $\partial_\mu \frac{\pi^i}{\chi_\pi}$ in Eqn.~(\ref{EOM1}). Then,
\bea \label{AbotEOM}
\partial_z \frac{e^{-\Phi(z)}}{z}\partial_z A_\mu^{i\bot}-\frac{e^{-\Phi(z)}}{z}\Box A_\mu^{i\bot} -\frac{2g_5^2 f^2(z)R^2}{k_s}\frac{e^{-\Phi(z)}}{z^3}A_\mu^{i\bot} =0,\\
\partial_z \frac{e^{-\Phi(z)}}{z}\partial_z A_\mu^{i\parallel}-\frac{2g_5^2 f^2(z)R^2}{k_s}\frac{e^{-\Phi(z)}}{z^3}A_\mu^{i\parallel}=0. \label{AparalEOM}
\eea

At the boundary we have the following terms in the effective $4D$ action:
\begin{align}\label{gen_BrBboundary}
\int d^4x & \left.\bigg[ e^{-\Phi(z)} \frac Rz \frac1{2g_5^2} A^{i\mu} (\partial_z A^i_\mu- \partial_\mu A^i_z)+
e^{-\Phi(z)} \frac {f^2(z)R^2}{z^3} \frac R{k_s}\frac{\pi^i}{\chi_\pi}\left(A^i_z-\partial_z \frac{\pi^i}{\chi_\pi}\right)\bigg]\right|_0^\infty\\
&\xrightarrow{\xi=1} - \frac1{2g_5^2} \int d^4x \left. \frac Rz  \left(A^{i\bot\mu} \partial_z A^{i\bot}_\mu +A^{i\parallel\mu} \partial_z A^{i\parallel}_\mu- A^{i\mu} \partial_\mu \partial_z \frac{\pi^i}{\chi_\pi} \right)\right|_{z=\varepsilon}\label{boundary_br}
\end{align}
The two-point function of the longitudinal mode is non-zero and that is the crucial difference from the previous sector.
The choice $\xi=1$ is explained in a  minute. For now we observe that it makes identical the bulk EOMs for $\pi^i$ and $ A_\mu^{i\parallel}$ and eliminates the Goldstone mass term from the boundary:
for $\xi=1$ all the Goldstones (including the component associated to the Higgs) are massless. It is also instructive 
to justify the system of EOMs (\ref{pionEOM})--(\ref{AparalEOM}) by deriving them in the model where $\xi=1$ is set 
from the start in Eqn.~(\ref{Az_of_xi}). That exercise is worked out in Appendix~\ref{xi1EOM}.

As in the unbroken case we perform the 4D Fourier transformation and 
establish the propagation between the source and the bulk for the transverse solution
\be
A_\mu^\bot(q,z)=\phi_{\mu}^{i\bot}(q)\cdot A(q,z),\qquad A(q,\varepsilon)=1.
\ee
Changing variables to $y=\kappa^2z^2$ we arrive at the following EOM
\be
y A''(q,y)-yA'(q,y)+\left(\frac{q^2}{4\kappa^2}-\frac{g_5^2(f(y)R)^2}{2yk_s}\right)A(q,y)=0.
\ee
An analytical solution of this EOM exists either for $f^2(y)\sim y$ or $f^2(y)\sim$ {const}. 
The last option taken together with the boundary condition 
on $A(q,z)$ leads to the implausible conclusion: $f(y)=0$. Therefore we turn to the linear ansatz
\be
f(z)=f\cdot\kappa z,
\ee
where the constant $f$ has the dimension of mass. We also introduce a convenient parameter 
\be
a=\frac{g_5^2(fR)^2}{2k_s}.
\ee
The bulk-to-boundary mode of the confluent hypergeometric equation above is specified as
\be\label{axprop}
A(q,\kappa^2z^2)=\Gamma\left(-\frac{q^2}{4\kappa^2}+1+a\right)\Psi\left(-\frac{q^2}{4\kappa^2}+a,0;\kappa^2z^2\right). 
\ee
The other mode for discrete values of $q^2$ and
$\left.A(q,z)\right|_{q^2=M_A^2(n)}=A_n(z)$ gives the $z$-profiles and masses of the eigenstates
\be\label{amodes}
A_n (z)=\kappa^2z^2 \sqrt{\frac{g_5^2}R}\sqrt\frac2{n+1} L_n^1(\kappa^2z^2),\quad M_A^2(n)=4\kappa^2\left(n+1+a\right),
\quad  n=0,1,2....
\ee
The orthogonality relation is completely equivalent to that of Eqn.~(\ref{unbr_orthog}). In fact, the only difference 
is that the intercept of the Regge trajectory is larger than in the unbroken case, though the pattern is identical. 
These states are heavier than their unbroken counterparts just as in QCD axial vector mesons are heavier than 
the vector ones.

The bulk-to-boundary solution of the longitudinal EOM~(\ref{AparalEOM}) is
\be
A_\mu^{i\parallel}(q,z)=\phi_\mu^{i\parallel}(q)\cdot A^{\parallel}(q,z),\ A^\parallel(q, z)=\Gamma\left(1+a\right)\Psi\left(a,0;\kappa^2z^2\right).
\ee
That is equivalent to the transverse propagator of Eqn.~(\ref{axprop}) but with $q^2=0$.

The Goldstone EOM~(\ref{pionEOM}) is the same as Eqn.~(\ref{AparalEOM}). However, $\frac12 (\partial_\mu \pi^i(x))^2$ 
is the correct normalization of the  Goldstone kinetic term in the 4D effective Lagrangian appearing after the 
integration over the $z$-dimension, and that fixes the constant factor differently
\be \label{pion_sol}
\pi(x,z)=F^{-1}\chi_\pi \Gamma\left(1+ a\right) \Psi\left(a,0;\kappa^2z^2\right),
\ee
where 
\be\label{FinEOM}
F^2=-\frac{2R\kappa^2a}{g_5^2} \left(\ln\kappa^2\varepsilon^2 +2\gamma_E+\psi\left(1+a\right)\right).\ee
In Section~\ref{sec-corr-vector} we will find the same $F^2$ in the residue of the massless pole of the broken 
vector correlator. The exact accordance is only possible for $\xi=1$. Furthermore, solution~(\ref{pion_sol}) fixes the due boundary interaction
\be \label{phi_pi_mixing}
\int d^4x (-F) \partial^\mu \pi^i(x)\phi^i_\mu(x).
\ee

As a result of $W^\alpha_\mu$ and $B_\mu$ couplings in Eqn.(\ref{lagr1}) the mixing in Eqn.(\ref{phi_pi_mixing}) 
for $i=1,2,3$ implies that the three Goldstones would be eaten by the SM gauge bosons to provide them masses 
proportional to $F$. Notice that there is no physical source to mix with the fourth Goldstone, it remains 
in the model as the physical Higgs particle $\pi^4(x)=h(x)$.
The phenomenological discussion of its properties  are postponed to a latter section.

To end the section, we introduce a convenient expression for the bulk-to-boundary propagators as the sums over 
the resonances (one should utilize Eqns.~(\ref{Tricrel}) and (\ref{app:inf_series_Tricomi}): 
\bea\label{D_btbprop2}
V(q,z)=\sum\limits\limits_{n=0}^\infty\frac{F_V(n)V_n(z)}{-q^2+M^2_V(n)},
\quad
A(q,z)=\sum\limits\limits_{n=0}^\infty\frac{F_A(n) A_n(z)}{-q^2+M^2_A(n)},\\ 
F_A^2(n)=F_V^2(n)=\frac{8R\kappa^4}{g_5^2}(n+1).
\eea
Here $F_{V/A}(n)$ are the decay constants related to the states with the corresponding quantum numbers.
The longitudinal broken and Goldstone solutions could be represented by infinite sums too.


\section{Two--point correlation functions}
\label{sec-corr}
\subsection{Unbroken and broken correlators} \label{sec-corr-vector}
The holographic prescriptions given in Eqns. (\ref{Zqft}) and (\ref{cor_corresp}) 
allow us to define the correlation function as
\be
\langle \mathcal O_\mu^{a/i}(q) \mathcal O_\nu^{b/j}(p)\rangle=\delta(p+q)\int d^4x e^{iqx}\langle \mathcal O_\mu^{a/i}(x) \mathcal O_\nu^{b/j}(0)\rangle
=\frac{\delta^2 iS_{boundary}^{5D}}{\delta i \phi^{a/i}_{\mu}(q)\delta i \phi^{b/j}_{\nu}(p)},
\ee
where the boundary remainders of the on-shell action were established in Eqns.~(\ref{boundary_unbr}) 
and (\ref{boundary_br}). We further define the  correlators
\begin{eqnarray}
	&i\int d^4x e^{iqx}\langle\mathcal O_\mu^{a/i}(x)\mathcal O_\nu^{b/j}(0)\rangle_\bot=\delta^{ab/ij}\left(\frac{q_\mu q_\nu}{q^2}-\eta_{\mu\nu}\right)\Pi_{unbr/br}(q^2),\\
&i\int d^4x e^{iqx}\langle\mathcal O_\mu^{i}(x)\mathcal O_\nu^{j}(0)\rangle_\parallel=\delta^{ij}\frac{q_\mu q_\nu}{q^2}\Pi^\parallel_{br}(q^2).
\end{eqnarray}
We should take into account that $\Pi_{unbr/br}(q^2)$ are subject to short distance ambiguities
of the form $C_0+C_1 q^2$ (see {\it e.g.} Refs.~\cite{REINDERS1985, Afonin2006}).
 
Performing the due variation in Eqn.~(\ref{boundary_unbr}) we find $\Pi_{unbr}(q^2)$ to be 
\be\label{unbrA}
\Pi_{unbr}(q^2)=\frac{R}{g_5^2}\left.\left[\frac{e^{-\Phi(z)}V(q,z)\partial_z V(q,z)}{z}\right]\right|_{z=\varepsilon}.
\ee
Let us substitute the propagator from Eqn.~(\ref{vprop}), then
\be\label{unbrA1}
\Pi_{unbr}(q^2)=-\frac{R}{2g_5^2} q^2 \left(\ln \kappa^2\varepsilon^2+2\gamma_E+\psi\left(-\frac{q^2}{4\kappa^2}
+1\right)\right),
\ee
where $\gamma_E$ is the Euler-Mascheroni constant and $\psi$ is the digamma function.

To separate the short distance ambiguities  we perform a decomposition of the digamma function 
(see Eqn.~(\ref{app:digamma_series})) in Eqn.~(\ref{unbrA1})
\be\label{unbrA2}
\Pi_{unbr}(q^2)=-\frac{R}{2g_5^2}  \left(\ln \kappa^2\varepsilon^2+\gamma_E\right)q^2
-\frac{2\kappa^2R}{g_5^2}  \sum\limits_{n=0}^\infty\frac{q^4}{M^2_V(n)(q^2-M^2_V(n))}.\ee
The first term would correspond to the ambiguity parametrizing constant $C_1$, while the second one is a well convergent sum over the
resonances.

An alternative procedure, introducing the resonances at an earlier stage with the use of the bulk-to-boundary 
propagator~(\ref{D_btbprop2}), should result in the same two-point function. Taking into account the 
orthogonality relation~(\ref{unbr_orthog}) we get from Eqn.~(\ref{unbrA})
\be
\Pi_{unbr}(q^2)=\sum\limits_{n=0}^\infty\frac{F^2_V(n)}{-q^2+M^2_V(n)}.
\ee
The ambiguities appear as follows
\be \label{unbrA3}
\Pi_{unbr}(q^2)=-\frac{2\kappa^2R}{g_5^2} \sum\limits_n\frac{q^4}{M^2_V(n)(q^2-M^2_V(n))}
+q^2\sum\limits_n\frac{2\kappa^2R/g_5^2}{M^2_V(n)}+\sum\limits_n\frac{2\kappa^2R}{g_5^2}.
\ee
After the proper subtractions, we are left with the first sum of Eqn.~(\ref{unbrA3}). This is the part 
relevant  for the resonance description of the two-point function that coincides with the sum in 
Eqn.~(\ref{unbrA2}). Hence, the convergent correlator is 
\be\label{Piunbr_subtracted}
\widehat\Pi_{unbr}(q^2)=\sum\limits_{n=0}^\infty\frac{q^4F_V^2(n)}{M^4_V(n)(-q^2+M^2_V(n))}.
\ee
Concerning the subtractions, it is not surprising that they differ for the correlators derived in two different 
ways. It is fundamental that they are limited to the form $C_0 +C_1 q^2$, but any reordering of the manipulations 
may affect the results as this is a divergent and ill-defined at short distances quantity. 
However, it is interesting to match the two expressions of $C_1$ in the proportional to $q^2$ terms of 
Eqns.~(\ref{unbrA3}) and (\ref{unbrA2}). To do that we need to introduce a regulator in the ``resonance'' 
representation -- a finite number of terms in the sum, a bound at some $N_{max}$.
Then a connection between the maximum number of resonances $N_{max}$ and the UV regulator $\varepsilon$ is
\be \label{regulator_connection}
\log N_{max}=-2\gamma_E-\log \kappa^2\varepsilon^2.
\ee
This relation is meaningful only at the leading order ({\it i.e.} the constant non-logarithmic part cannot be
 determined by this type of heuristic arguments).
Finally, the last sum in Eqn.~(\ref{unbrA3}) behaves as $\sim N_{max}^2$ if we sum up a finite number of 
resonances and actually corresponds to a potentially  subleading logarithmic divergence. Therefore, it can be 
eliminated by setting the subtraction constant $C_0$.

In the broken vector sector the situation is very similar. For the transverse modes, variation of 
Eqn.~(\ref{boundary_br}) results in 
\be
\Pi_{br}(q^2)=\frac{R}{g_5^2}\left.\left[\frac{e^{-\Phi(z)}A(q,z)\partial_z A(q,z)}{z}\right]\right|_{z=\varepsilon}.
\ee
Substituting the propagator from Eqn.~(\ref{axprop}) leads to
\be \label{brA}
\Pi_{br}(q^2)=-\frac{R}{2g_5^2} q^2 \left(1-\frac{4\kappa^2a}{q^2}\right)\left(\ln \kappa^2\varepsilon^2
+2\gamma_E+\psi\left(-\frac{q^2}{4\kappa^2}+1
+a\right)\right).
\ee
An alternative expression for the two-point correlator is
\be
\Pi_{br}(q^2)=\sum\limits_{n=0}^\infty\frac{F^2_A(n)}{-q^2+M^2_V(n)}.
\ee 
And in both cases the subtraction of short distance ambiguities leads to
\be
\widehat\Pi_{br}(q^2)=\sum\limits_n\frac{q^4 F_A^2(n)}{M^4_A(n)(-q^2+M^2_A(n))}-F^2, \label{Pibr_subtracted}
\ee
where we find a ``pion'' pole with  the ``pion decay constant'' $F$ anticipated in Eqn.~(\ref{FinEOM}) and derived 
there from a completely different argument. It could also be expressed in the form of an infinite series
\be
F^2=\frac{2R\kappa^2a}{g_5^2}\sum\limits_n \frac{1}{n+1+a}. \label{FinSum}
\ee
Variation over the longitudinal modes in Eqn.~(\ref{boundary_br}) also brings this constant
\be
\Pi^\parallel_{br}(q^2)=F^2.
\ee

Once more, fulfilling relation~(\ref{regulator_connection}) makes an accordance between the order-$q^2$ 
subtractions. This demonstrates the ultraviolet origin of the renormalization ambiguity involved in the 
constant $C_1$ because the outcome is independent on whether we treat the broken or unbroken symmetries. 
The same could be implied about $C_0$. Then, the determination of $F^2$ in (\ref{FinSum}) is straightforward 
as soon as we subtract the ``quadratic'' term $\sum\limits_n \frac{2\kappa^2R}{g_5^2}$.

In the end, these correlation functions appear in the 4D effective Lagrangian as 
\be
\mathcal L_{eff} \supset \frac12 \phi_\mu^a \left(\frac{q_\mu q_\nu}{q^2}-\eta_{\mu\nu}\right)\Pi_{unbr} \phi_\nu^a 
+\frac12 \phi_\mu^i\left(\left(\frac{q_\mu q_\nu}{q^2}-\eta_{\mu\nu}\right)\Pi_{br}+\frac{F^2 q^\mu q^\nu}{q^2}\right) \phi_\nu^i.
\ee

\subsection{Vacuum polarization amplitudes of the gauge fields}
We started the discussion about the holographic CH model assuming that the SM gauge fields couple to the 
currents of the strongly interacting sector $\widetilde J^{\alpha}_{L\mu}$ and $\widetilde J^{3}_{R\mu}$ as 
in Eqn.~(\ref{lagr1}). These currents are proportional to the ones dual to the $5D$ fields, $\mathcal O^{a/i}_\mu$, 
with the EW couplings $g$ and $g'$ necessarily appearing. We introduced the factor $g_V$ to modulate that 
proportionality. The misalignment should also be taken into account. 
In the notation of Eqn.~(\ref{rotated_gen}), a rotated operator can be given in terms of the original 
ones as ($\alpha,i=1,2,3$ here)
\be
\widetilde{\mathcal O}^{\alpha}_{L/R\mu}=\frac{1\pm\cos\theta}{2}
\mathcal O^{\alpha}_{L\mu} +\frac{1\mp\cos\theta}{2}
\mathcal O^{\alpha}_{R\mu} \mp \frac{\sin\theta}{\sqrt2}\mathcal O^i_{\mu}.
\ee

The two-point correlators of physical interest are
\begin{align}
i\int d^4xe^{iqx} \langle \widetilde J^{\alpha}_{L\mu}(x) \widetilde J^{\beta}_{L\nu}(0)\rangle&=\delta^{\alpha\beta}\frac{g^2}2
\left[\left(\frac{q_\mu q_\nu}{q^2}-\eta_{\mu\nu}\right)\Pi_{LL}(q^2)+\frac{q_\mu q_\nu}{q^2}\Pi_{LL}^{\parallel}(q^2)\right],\\
i\int d^4xe^{iqx} \langle \widetilde J^{\alpha}_{R\mu}(x) \widetilde J^{\beta}_{R\nu}(0)\rangle&=\delta^{\alpha\beta}\frac{g'^2}2 
\left[\left(\frac{q_\mu q_\nu}{q^2}-\eta_{\mu\nu}\right)\Pi_{RR}(q^2)+\frac{q_\mu q_\nu}{q^2}\Pi_{RR}^{\parallel}(q^2)\right],\\ \label{LRdef}
2i\int d^4xe^{iqx} \langle\widetilde J^{\alpha}_{L\mu}(x) \widetilde J^{\beta}_{R\nu}(0)\rangle&=
\delta^{\alpha\beta}\frac{gg'}2\left[\left(\frac{q_\mu q_\nu}{q^2}-\eta_{\mu\nu}\right)\Pi_{LR}(q^2)+\frac{q_\mu q_\nu}{q^2}\Pi_{LR}^{\parallel}(q^2)\right];
\end{align}
where we defined the quantities
\begin{align}
&\Pi_{diag}(q^2)=\Pi_{LL}(q^2)=\Pi_{RR}(q^2)=\frac{1+\cos^2\theta}{2g_V^2}\Pi_{unbr}(q^2)+\frac{\sin^2\theta}{2g_V^2}\Pi_{br}(q^2),\\ \label{PiLR_def}
&\Pi_{LR}(q^2)= \frac {\sin^2\theta}{g_V^2}\left(\Pi_{unbr}(q^2)-\Pi_{br}(q^2)\right),\\
&\Pi_{LL}^{\parallel}(q^2)=\Pi_{RR}^{\parallel}(q^2)=\frac{\sin^2\theta}{2g_V^2} F^2,\ \Pi_{LR}^{\parallel}(q^2)= -\frac{\sin^2\theta}{g_V^2}F^2.
\end{align}

The relevant quadratic contribution of the gauge bosons to the 4D partition function is
\begin{align}
\notag
\mathcal L_{eff} \supset&\left(\frac{q^\mu q^\nu}{q^2}-
\eta^{\mu\nu}\right)\frac{1}{4}\Pi_{diag}(q^2)(g^2{W}_\mu^{\alpha}{W}_\nu^{\alpha}+g'^2{B}_\mu{B}_\nu)
\\ 
&+\frac{ F^2\sin^2\theta}{8g_V^2}\frac{q^\mu q^\nu}{q^2}(g^2{W}_\mu^{\alpha}{W}_\nu^{\alpha}+g'^2{B}_\mu{B}_\nu)&  \label{gaugedL}\\ 
&+\left(\frac{q^\mu q^\nu}{q^2}-\eta^{\mu\nu}\right)\frac{1}4 \Pi_{LR}(q^2) gg'{W}_\mu^{3}{B}_\nu
-\frac{q^\mu q^\nu}{q^2}\frac{F^2\sin^2\theta}{4g_V^2} gg'{W}_\mu^{3}{B}_\nu\nn.
\end{align}
The mass terms in the effective Lagrangian can be determined from the lowest order in $q^2$. Both for 
the longitudinal and transverse $W$ and $Z$ gauge bosons we get
\be
M^2_W=\frac{g^2}{4} \frac {\sin^2\theta}{g_V^2}  F^2,\ M^2_Z=\frac{g^2+g'^2}4\frac {\sin^2\theta}{g_V^2} F^2, \label{Gauge_masses}
\ee
while the photon stays masless.

\subsection{Left--right correlator and  sum rules}
The vacuum polarization amplitudes receive contributions from the new physics (new massive resonances in the loops). 
To quantify deviations with respect to SM, the EW ``oblique'' precision parameters were introduced \cite{AltarelliB, PT}.  The most relevant for the discussion of the CH models are the $S$ and $T$ parameters of 
Peskin and Takeuchi~\cite{PT}. As we already mentioned, a particular feature of MCHM is that due to the 
custodial symmetry of the strongly interacting sector the tree-level correction to the $T$ parameter vanishes. 
Bearing in mind that the holographic description is meant to be valid only in the large $N_{hc}$ limit, loop corrections
are not easily tractable. Thus, we focus on the $S$ parameter connected to the $\Pi_{LR}(q^2)$ as follows
\be\label{Spar1}
S=-4\pi\Pi_{LR}'(0)=\frac{2\pi R}{g_5^2} \frac{\sin^2\theta}{g_V^2} \left[\gamma_E+\psi\left(1+a\right)+a\psi_1\left(1+a\right) \right].
\ee
Alternatively, it could be expressed through masses and decay constants:
\be \label{Spar2}
S=4\pi\frac{\sin^2\theta}{g_V^2} \bigg[\sum\limits_n\frac{F^2_V(n)}{M^4_V(n)}-\sum\limits_n\frac{F^2_A(n)}{M^4_A(n)}\bigg].
\ee
The experimental bounds on the $S$ parameter are essential for the numerical analysis of Section~\ref{phen}.
 
Further, we would like to investigate the validity of the equivalent of the Weinberg sum rules (WSR) that
relate  the imaginary part of $\Pi_{LR}(q^2)$ to masses and decay constants of vector resonances
in the broken and unbroken channels, respectively. 
We start with the subtracted correlators $\widehat \Pi_{unbr}$ and $\widehat \Pi_{br}$ of 
Eqns. (\ref{Piunbr_subtracted}) and (\ref{Pibr_subtracted}), then select a suitable integration circuit 
and formally obtain 
\begin{align}\label{Vint}
\frac1\pi\int_0^\infty \frac{dt}{t}  \text{Im} \Pi_{unbr}(t)&=\sum\limits_n \frac{F_V^2(n)}{M^2_V(n)},\\ \label{Aint}
\frac1\pi\int_0^\infty \frac{dt}{t}  \text{Im} \Pi_{br}(t)&=\sum\limits_n\frac{F_A^2(n)}{M^2_A(n)}+F^2.
\end{align}
However, these expressions are ill-defined: the external contour does not vanish, and the imaginary 
part of the poles should have been specified. The latter can be done following Vainshtein, {\it i.e.} 
replacing $M_V^2 (n)$ in Eqn.~(\ref{Piunbr_subtracted}) with $M_V^2 (n)(1 - i\epsilon)$. This prescription 
reproduces the correct residues.
Additionally, the left hand sides are generically divergent while the sum over resonances possesses an 
essential singularity on the real axis when the number of resonances $N_{max}$ encircled in the contour 
tends to infinity. 

We expect to see the convergence properties of the integrals on the left hand side of (\ref{Vint}) and 
(\ref{Aint}) improved when they are gathered in the left-right combination. For the uniformity of notation 
we introduce the sum  $F^2=\sum\limits_{n<N_{max}}F^2(n)$ (from Eqn. (\ref{FinSum})). Then,
\be\label{WSR1}
\frac{1}{\pi}\int_0^{M^2(N_{max})} \frac{dt}{t} \text{Im} \Pi_{LR}(t)
=\frac{\sin^2\theta}{g^2_V} \sum\limits_{n<N_{max}}\left(\frac{F_V^2(n)}{M^2_V(n)} - \frac{F_A^2(n)}{M^2_A(n)}- F^2(n)\right).
\ee
In QCD $\Pi_{LR}$ decays fast enough so that the external contour contribution is negligible when 
enough resonances are encircled, and this integral vanishes. 
The equality of Eqn.~(\ref{WSR1}) to zero is the first WSR for QCD, and the same arguments allow one to 
derive the second WSR
\be\label{WSR2}
\frac{1}{\pi}\int_0^{M^2(N_{max})} dt \text{Im} \Pi_{LR}(t)
=\frac{\sin^2\theta}{g^2_V}\sum\limits_{n<N_{max}}(F_V^2(n) - F_A^2(n))=0.
\ee
In fact, it is well known that in QCD including just the first resonances in the sum provides a fair 
agreement with phenomenology~\cite{Weinberg67}.
In any case, the convergence of the dispersion relation (no subtraction is needed) indicates that 
the limit $N_{max}\to \infty$ could be taken in QCD.

To understand whether the  situation is indeed analogous to QCD 
we should address these 
two questions: (a) can the contour integral be neglected? (b) if so, is the integral on the left hand side converging?

To answer the first question we consider $\Pi_{LR}(Q^2)$, given explictly in Eqn.~(\ref{LR0}) with Euclidean momenta 
$Q^2=-q^2$, and expand it for large $Q^2$ (we make use of the Stirling's expansion of the $\psi$ function)
\be
\frac{g^2_V \Pi_{LR}(Q^2)}{Q^2}=\sin^2\theta\frac{2\kappa^2 a}{Q^2}\frac R{g_5^2}\left(\ln \frac{Q^2}{4\kappa^2}
+\ln\kappa^2\varepsilon^2-\frac{2\kappa^2a}{Q^2}\right)+\mathcal O\left(\frac1{Q^6}\right).
\ee
This limit is constrained to the (unphysical) region of $|\arg Q^2|<\pi$, while the value on the physical 
axis ($0 < \text{Re}\ q^2=-\text{Re}\ Q^2$) stays ill-defined (needs a prescription, such as the one discussed above). 
However, we are now in position to discuss the convergence of the outer part of the circuit in  Eqns.~(\ref{WSR1}) 
and (\ref{WSR2}). Due to the presence of the $\ln Q^2/Q^2 $ and $1/Q^2$ terms the correlator does not vanish 
fast enough to make the issue similar to the QCD case.
Therefore, the corresponding dispersion relation requires one subtraction constant $c$ to parametrize the 
part of $\Pi_{LR}(Q^2)$ not determined by its imaginary component
\be
\frac{\Pi_{LR}(Q^2)}{Q^2}=\int_0^\infty\frac{dt}{t+Q^2-i\epsilon} \frac1\pi\frac{\text{Im} \Pi_{LR}(t)}{t}+c.
\ee 

In the deep Euclidean region one could use an expansion
\be\label{LR2}
\frac{1}{t+Q^2}=\frac1{Q^2}-\frac1{Q^2}t\frac1{Q^2}+...\ee
and then the dispersion relation in the large $Q^2$ limit looks as
\be \label{LR3}
\frac{\Pi_{LR}(Q^2)}{Q^2}=c+ \frac1{Q^2}\frac1\pi\int_0^\infty \frac{dt}{t}  \text{Im}\Pi_{LR}(t)-
\frac1{Q^4}\frac1\pi\int_0^\infty dt 
 \text{Im}\Pi_{LR}(t)+ \ldots 
\ee

The next step is to encircle a large, but finite, number of resonances. That is, we take  $N_{max} <\infty$ 
connected to the UV cut-off via the relation~(\ref{regulator_connection}). The dispersion relation  
still holds and Eqn.~(\ref{LR3}) can be compared order by order with the large $Q^2$ expansion given in 
Appendix~\ref{Q2exp}.  Holding to the assumptions made there, we obtain 
\be 
\int_0^{M^2(N_{max})} \frac{dt}{t}  \text{Im}\Pi_{LR}(t) =0,
\ee
that establishes the formal validity of the first WSR
\be
 \sum\limits_{n<N_{max}}\left(\frac{F_V^2(n)}{M^2_V(n)} - \frac{F_A^2(n)}{M^2_A(n)}- F^2(n)\right) = 0.
\ee
We further stress that the situation is rather unsimilar to the one of real QCD, essentially because $F^2$ 
is logarithmically dependent on the cut-off. On the other hand, the situation in the holographic CH scenario 
is quite analogous to the holographic QCD model of Ref.~\cite{Espriu_2020}. We just proved 
that the sum over vector resonances  
$\sum\limits_{n<N_{max}}\left(\frac{F_V^2(n)}{M^2_V(n)} - \frac{F_A^2(n)}{M^2_A(n)}\right)$ is itself cut-off 
dependent for $N_{max} \to \infty$. This implies that symmetry restoration 
takes place very slowly in the UV and saturation with the ground state resonance is questionable both 
in holographic CH and holographic QCD. It seems fair to conclude that these peculiarities represent a 
pitfall of holography rather than a characteristic of the CH model.

Finally, the nullification of the $\frac1{Q^4}$ term in (\ref{largeQ2-2}) leads to
\be
\frac1\pi\int_0^{M^2(N_{max})} dt\,  \text{Im}\Pi_{LR}(t)=0,
\ee
that formally proves the second WSR of Eqn.~(\ref{WSR2}). Again, a cut-off should be imposed to guarantee 
convergence of both the integral of the imaginary part over the real axis and of the sum over resonances.

\section{Higher order correlators and couplings} \label{coupl_section}
Let us write down several 5D interactions of phenomenological interest. At the three-point level they are
\bea\label{triple_int}
S^{(3)}_{5D}\supset&i\frac R{g_5^2}\int d^5x e^{-\kappa^2z^2} z^{-1}\left(\partial_\mu A_\nu^A A^{B \mu} A^{C\nu} \Tr T^A[T^B,T^C] 
- \partial_z A_\mu^A A_z^i A^{B \mu} \Tr T^A[T^i,T^B]\right.\\
&\left.+\partial_\mu A_z^i A_z^j A^{A\mu}\Tr T^i[T^j,T^A]\right)
+(fR)^2 \kappa^2\frac R{k_s}\int d^5xe^{-\kappa^2z^2} z^{-1} \frac{h}{\chi_\pi}(A_L-A_R)^\alpha_\mu A^{\alpha\mu}_{br}.\nn
\eea
To prevent misunderstanding we specify the left, right or broken origin of vector field $A_\mu(x,z)$ where 
it is needed (they go with $\alpha=1,2,3$). Otherwise, the fields with $i,j=1,2,3,4$ are from the broken sector, 
and $A,B,C=1,\ldots,10$ ones encompass all options. The fourth Goldstone field $\pi^4(x,z)$ is denoted as $h(x,z)$ 
henceforth. At the four-point level we have
\bea\nn
S^{(4)}_{5D}\supset&\frac R{4g_5^2}\int d^5x e^{-\kappa^2z^2} z^{-1}\left(A^A_\mu A^B_\nu A^{C\mu} A^{D\nu} \Tr[T^A,T^B] [T^C, T^D ]-2A_z^iA^A_\mu A^j_z A^{B\mu}\Tr[T^i,T^A] [T^j,T^B ]\right)\\
&+(fR)^2 \kappa^2\frac R{4k_s}\int d^5x\frac{e^{-\kappa^2z^2}}{z}\frac{h^2}{\chi_\pi^2}\left((A^\alpha_{L\mu}-A^\alpha_{R\mu})^2 - 2A^{\alpha2}_{br\mu}\right).
\eea
The commutators there can be simplified with the Lie algebra of $SO(5)$
\bea\nn
&[T^\alpha_L,T^\beta_L]=i\varepsilon^{\alpha\beta\delta}T^\delta_L,\ 
[T^\alpha_R,T^\beta_R]=i\varepsilon^{\alpha\beta\delta}T^\delta_R,\ 
[T^\alpha_L,T^\beta_R]=0,\ \alpha,\beta,\delta=1,2,3\\ \nn
&[T^a,\widehat{T}^i]=\widehat{T}^j(t^a)^{ji},\ [\widehat{T}^i,\widehat{T}^j]=(t_a)^{ji}T^a,\ a=1,\ldots,6,\ i=1,\ldots,4.\eea
Here $t^a=\{t^\alpha_L, t^\alpha_R\}$, see the definition after Eqn.~(\ref{genso4}).

The expressions for $S^{(3)}_{5D}$ and $S^{(4)}_{5D}$ are already simplified with the gauge choice $A_z^a=0$
in the unbroken channel. The Higgs-related terms proportional to $(fR)^2$ come from the square of the covariant derivative in Eqn.~(\ref{5Daction}). Taking into account that in the broken sector imposed with $\xi=1$ we 
had $A_z^i=\frac{\partial_z \pi^i}{\chi_\pi}$, we reveal the following interactions involving the Higgs from  
the $F_{MN}^2$ term
\be
\frac R{2g_5^2}  \int d^5x\frac{e^{-\kappa^2z^2}}{z}\left[\frac{\partial_z h}{\chi_\pi}(A_L-A_R)^\alpha_\mu \partial_z A^{\alpha\mu}_{br} +\frac14 \left(\frac{\partial_z h}{\chi_\pi}\right)^2\left((A^\alpha_{L\mu}-A^\alpha_{R\mu})^2 +A^{\alpha2}_{br\mu}\right)\right].
\ee

We are interested in triple and quartic couplings between the Higgs boson and the SM gauge bosons. 
In the standard  MCHM picture these interactions have a given parametrization in the coordinate space
\bea
&g^{SM}_{hWW}\cos\theta W^+_\mu W^{-\mu}h+g^{SM}_{hZZ}\cos\theta\frac12 Z_\mu Z^{\mu}h+ \frac{\cos2\theta}{4}\left(g^2W^+_\mu W^{-\mu}+\frac{g^2+g'^2}{2}Z_\mu Z^\mu\right)hh,\label{MCHM_higgs_inter}\\
	&g^{SM}_{hWW}=g M_W,\ g^{SM}_{hZZ}=\sqrt{g^2+g'^2} M_Z, \label{SM_higgs_coupl}
\eea
with $W^{\pm}_{\mu}=\frac{W_\mu^1 \mp i W_\mu^2}{\sqrt 2},\ Z_\mu=\frac1{\sqrt{g^2+g'^2}}\left(gW_\mu^3-g' B_\mu\right)$.

In our 5D model the effective couplings for $hWW$ and $hhWW$ originate from
\begin{align}\label{part_hww}
\mathcal L_{eff}\supset &i\frac{g^2}{4g_V^2}h(q)W^{\alpha\ \mu}(k_1)W^{\beta\ \nu}(k_2) \langle h(q)|\widetilde{\mathcal O}^{\alpha}_{L\mu}(k_1)\widetilde{\mathcal O}^{\beta}_{ L\nu}(k_2)|0\rangle\\\label{part_hhww}
&+ i\frac{g^2}{4g_V^2}h(q_1)h(q_2)W^{\alpha\ \mu}(k_1)W^{\beta\ \nu}(k_2) \langle h(q_1)h(q_2)|\widetilde{\mathcal O}^{\alpha}_{L\mu}(k_1)\widetilde{\mathcal O}^{\beta}_{ L\nu}(k_2)|0\rangle.
\end{align}
$Z$ boson couplings can be taken into consideration after addition of the terms generated by 
$\widetilde{\mathcal O}^{3}_{L\mu}\widetilde{\mathcal O}^{3}_{ R\nu}$, 
$\widetilde{\mathcal O}^{3}_{R\mu}\widetilde{\mathcal O}^{3}_{ L\nu}$ and 
$\widetilde{\mathcal O}^{3}_{R\mu}\widetilde{\mathcal O}^{3}_{ R\nu}$ operator combinations. Their derivation 
follows closely that of the $W^+W^-$, so we just include them in the final result.

Particularities of calculating the matrix elements in (\ref{part_hww}) and (\ref{part_hhww}) can be found 
in Appendix~\ref{higgs_coupl_var}. The couplings to the EW gauge bosons appear in the effective Lagrangian as
\begin{align}
	\mathcal L_{eff} \supset 
	& \frac{g^{SM}_{hWW}\cos\theta}{\sqrt 2 g_V} \cdot  \frac12 \left(W^{+\mu}(k_2) W^{-}_{\mu}(k_1)+W^{-\mu}(k_2) W^{+}_{\mu}(k_1)\right) h(q)\\
	&+\frac{g^2 \cos2\theta}{8g_V^2}\cdot \frac12  \left( W^+_\mu(k_1) W^{-\mu}(k_2)+W^-_\mu(k_1) W^{+\mu}(k_2)\right) h(q_1)h(q_2)
	\\&+\frac{g^{SM}_{hZZ}\cos\theta}{\sqrt 2 g_V}\cdot \frac12  Z^{\mu}(k_2) Z_{\mu}(k_1) h(q) \\&+\frac{(g^2+g'^2)\cos2\theta}{8g_V^2}\cdot\frac{1}{2} Z_\mu(k_1) Z^{\mu}(k_2) h(q_1)h(q_2),\\
	&g^{SM}_{hWW}=\frac{g^2 F\sin\theta }{2 g_V},\ g^{SM}_{hZZ}=\frac{(g^2+g'^2) F\sin\theta}{2 g_V},
\end{align} 
where the factors in the last line indeed correspond to the SM notation of Eqn.~(\ref{SM_higgs_coupl}) due 
to the definition of masses in Eqn.~(\ref{Gauge_masses}).
The only thing missing to have the exact MCHM factors of Eqn.~(\ref{MCHM_higgs_inter}) is the proper choice 
of the so far free parameter
\be\label{gV_def}
g_V=\frac1{\sqrt2}.
\ee
Note, that this value is obtained in the approximation $M^2_W\ll 4\kappa^2$ assumed in the calculations of Appendix~\ref{higgs_coupl_var}.

Let us now turn to the part of Eqn.~(\ref{triple_int}) independent of $A_z$ and Higgs modes 
\be\label{triple_int_A}
i\frac R{g_5^2}\int d^5x e^{-\kappa^2z^2} z^{-1}\partial_\mu A_\nu^A A^{B \mu} A^{C\nu} \Tr T^A[T^B,T^C]
\ee
 The commutator is proportional to the epsilon-tensor if none of the three fields is $A_{br}^4$. 
In the oppsite case 
we rather obtain a Kronecker delta. 

There is an interaction between three vector 5D fields in Eqn.~(\ref{triple_int_A}). In order to procure 
a coupling of a vector resonance to two EW gauge bosons one of the fields should be taken in its KK 
representation, while the other two should be given by their bulk-to-boundary propagators coupled later 
to the corresponding gauge field sources. The details of these calculations are presented in 
Appendix~\ref{resonance_coupl_var}.

We limit ourselves to listing just the interactions for the ground states of the composite resonances
\bea \nn
\mathcal L_{eff}&\supset& \frac12 W^\alpha_{\mu_2}(q_2) W^\beta_{\mu_3}(q_3)\texttt{Lor}^{\mu_1\mu_2\mu_3}(q_1,q_2,q_3)(-i \varepsilon^{\alpha\beta\delta})\\ &\times&  \left( A_{\mu_1}^{L\ \delta}(q_1) g_{LWW}+A_{\mu_1}^{R\ \delta}(q_1) g_{RWW}-A_{\mu_1}^{Br\ \delta}(q_1) g_{BrWW} \right)\\
&+& W^\alpha_{\mu_2}(q_2) B_{\mu_3}(q_3)\texttt{Lor}^{\mu_1\mu_2\mu_3}(q_1,q_2,q_3)(-i \varepsilon^{\alpha3\delta})\\ &\times&  \left( A_{\mu_1}^{L\ \delta}(q_1) g_{LWB}+A_{\mu_1}^{R\ \delta}(q_1) g_{RWB}\right),
\eea
where the notation $\texttt{Lor}^{\mu_1\mu_2\mu_3}(q_1,q_2,q_3)$ was given in Appendix~\ref{resonance_coupl_var}, 
and we introduced
\bea \label{g_unbr_simpl}
&g_{L/RWW}=\frac{g^2}{4g_V^2} \sqrt{\frac{R}{2g_5^2}}\left[1\pm\cos\theta+ a\sin^2\theta(a\psi_1(1+a)-1)\right],\\ 
\label{g_br_simpl} &g_{BrWW}=\frac{g^2}{4g_V^2} \sqrt{\frac{R}{g_5^2}}\frac{\sin\theta}{1+a},\\
& g_{LWB}=g_{RWB}=\frac{gg'}{4g_V^2} \sqrt{\frac{R}{2g_5^2}}a\sin^2\theta\left[1-a\psi_1(1+a)\right]. \label{g_WB_simpl}
\eea
The numerical values of these couplings will be estimated in the next section.

\section{Numerical results for masses and couplings}\label{phen}
A very stringent limit on any new physics contribution comes from the experimental bounds on the $S$ parameter,
calculated using 5D techniques in Eqn.~(\ref{Spar1}) or (\ref{Spar2}). Recent EW precision data 
(see Ref.~\cite{PDG2020}) constraints it to the region
\be
S=-0.01\pm 0.10.
\ee
There are three a priori free parameters in our expression for $S$: $\sin\theta$, $a$ and $\frac{R}{g_5^2}$; 
and $g_V$ is assumed to be fixed as in Eqn.~(\ref{gV_def}). $a$ is related to the symmetry breaking by $f(z)$: 
at $a=0$ there is no breaking, the unbroken and broken vector modes have the same mass. 
In principle, $\frac{R}{g_5^2}$ could be evaluated by comparing holographic two-point function to the 
perturbative calculation of the Feynman diagram ({\it e.g.}, of a hyper-fermion loop) at the leading order 
in large $Q^2$ momenta, as it is usually done in the holographic realizations of QCD. As we would expect to get the 
hyper-color trace in the loop, it could be estimated that there is a proportionality $\frac{R}{g_5^2}\propto N_{hc}^p$ 
(power $p$ depends on the particular representation). However, we deliberately made no hypothesis on the 
fundamental substructure, and could only expect that very large values of $\frac{R}{g_5^2}$ correspond 
to the large-$N_{hc}$ limit. To have an idea of the scale of this quantity, we recall that for $N_c=3$ QCD 
one has $\frac{R}{g_5^2}\sim0.3$~\cite{Espriu_2020}.

\begin{figure}[t]
	\includegraphics[scale=0.45]{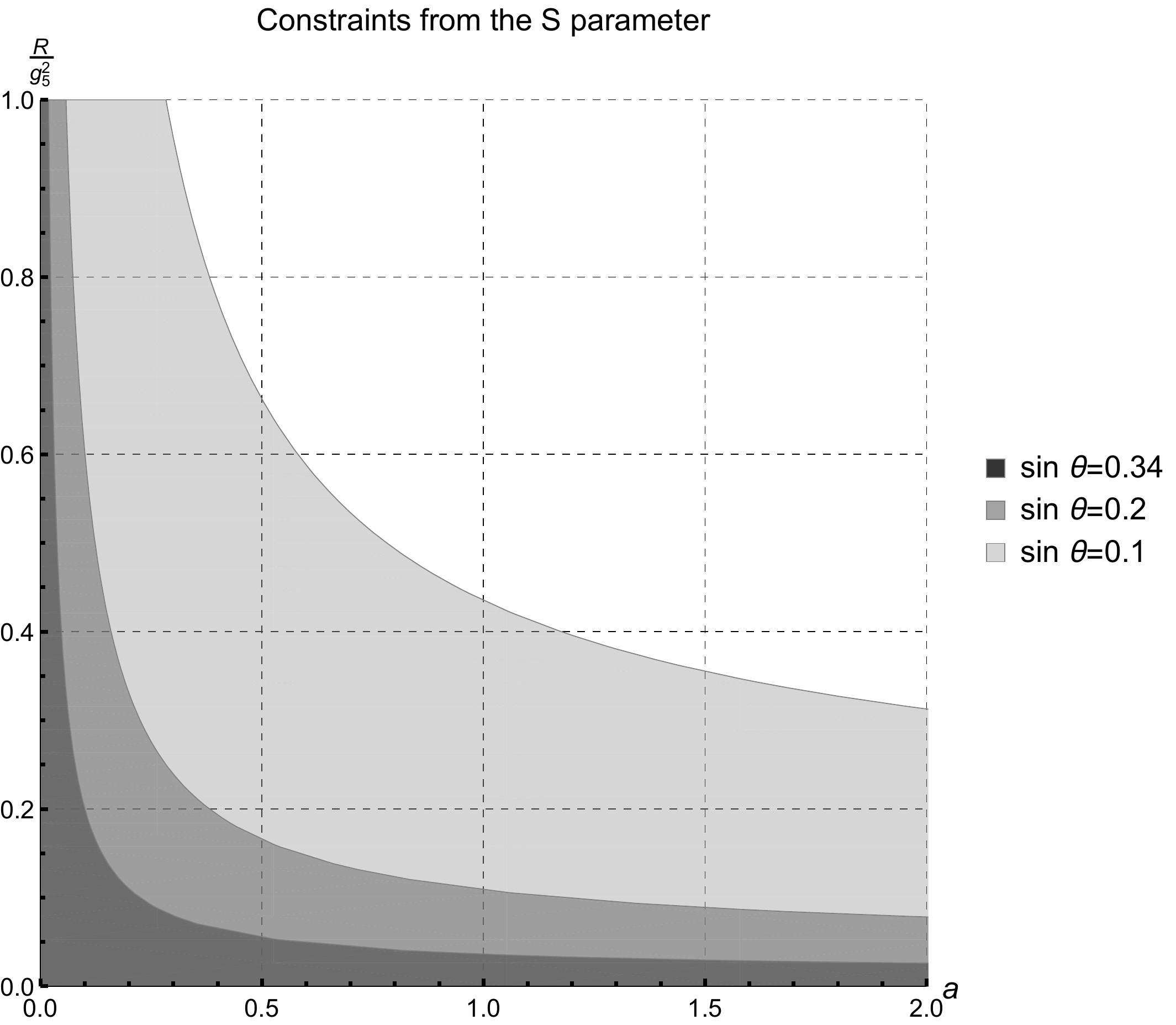}
	\caption{The $(\sin\theta, a, R/g_5^2)$ parameter region allowed by the $S$ parameter restraints. \label{Splot}
	}
\end{figure}

We present the effect of the current $S$-constraint on the $(\sin\theta, a, R/g_5^2)$ plane in  Fig.~\ref{Splot}.
The larger the value of $\sin \theta$ the smaller the allowed region for $a$ and $R/g_5^2$.
We only consider $\sin\theta\leq0.34$ due to the present bounds on the misalignment in MCHM~\cite{Atlas_2015} 
(for the the SM fermions in the spinoral representation of $SO(5)$). That bound is valid under the assumption 
that the coupling of the Higgs to gauge bosons is $\kappa_V=\sqrt{1-\sin^2\theta}$, and it was demonstrated 
in Section~\ref{coupl_section} that this is the case of our holographic model too. Otherwise, we 
can take a more model-independent estimation from the latest ATLAS and CMS combined measurements with the LHC 
Run 1 dataset~\cite{PDG2020} that lead to $\kappa_V=1.04\pm 0.05$ at one standard deviation. Taken at 
two standard deviations it results once again in $\sin\theta\leq0.34$. Nevertheless, stricter (lower) bounds
could  also be encountered in  the literature (Run 2 analyses~\cite{Aad2020,Sirunyan2019}, for instance).

No information on the mass scale $\kappa$ could be retrieved from the EW precision data.  However, we can 
relate it to the low-energy observables through the definition of the  $W$ boson mass in
Eqn.~(\ref{Gauge_masses}). It is connected to the EWSB scale $v=246$ GeV and we can equate
\be
M^2_W=\frac{g^2v^2}{4}=\frac{g^2F^2\sin^2\theta }{4 g_V^2}.
\ee
With $F$ given in Eqn.~(\ref{FinEOM}), the following condition on $\kappa$ is valid:
\be
\frac{g_V^2 v^2}{\sin^2\theta}+\frac{2\kappa^2R}{g_5^2}a \left(\ln\kappa^2\varepsilon^2 
+2\gamma_E+\psi\left(1+a\right)\right)=0.
\ee
Let us further set
\be\label{Mchar_cond}
\varepsilon=\frac1{\Lambda_\text{cut-off}}\simeq\frac1{4\pi f_{CH}}=\frac{\sin\theta}{4\pi v}.\ee
Here $\Lambda_\text{cut-off}=\Lambda_\text{CH}\simeq4\pi f_{CH}$ is the range of validity of the effective theory of the composite resonances, which could 
be postulated as a natural cut-off in the present bottom-up model. We can also rework the connection 
between the number of resonances cut-off $N_{max}$ and $\varepsilon$:
\be
N_{max}=16 \pi^2\frac{v^2}{\kappa^2\sin^2\theta} e^{-2\gamma_E}.
\ee

\begin{table}[t]
 \centering
\caption{Different predictions of the minimal vector masses for $\sin\theta=0.1,\ 0.2$ and $0.34$.}
\begin{tabular}{c | c | c | c | c | c}
$\ \sin\theta\ $ & $\ \frac{R}{g_5^2}\ $ & $\ a\ $  & $M_\ast=M_V(0)$, $\text{TeV}$ & $M_A(0)$, $\text{TeV}$ &$\sim N_{max}$\\
\hline
$0.1$& $0.1$ & $266.3$ & $0.22$ & $3.68$ & $>20\text{ k}$ \\
$0.1$& $0.3$ & $2.212$ & $1.28$ & $2.29$ & $740$ \\
$0.1$& $1$ & $0.283$ & $1.88$ & $2.13$ & $340$ \\
$0.1$& $10$ & $0.022$ & $2.10$ & $2.12$ & $270$ \\
\hline
$0.2$& $0.1$ & $1.176$ & $1.79$ & $2.64$ & $93$ \\
$0.2$& $0.3$ & $0.225$ & $2.28$ & $2.52$ & $58$ \\
$0.2$& $1$ & $0.058$ & $2.43$ & $2.50$ & $50$ \\
$0.2$& $10$ & $0.006$ & $2.49$ & $2.50$ & $48$ \\
\hline
$0.34$& $0.1$ & $0.225$ & $2.84$ & $3.14$ & $12$ \\
$0.34$& $0.3$ & $0.065$ & $3.00$ & $3.09$ & $11$ \\
$0.34$& $1$ & $0.019$ & $3.05$ & $3.08$ & $10$\\
$0.34$& $10$ & $0.002$ & $3.07$ & $3.08$ & $10$ \\
\end{tabular}
\label{Masses_table}
\end{table}

Setting $g_V=\frac 1{\sqrt2}$, we collect the results in Table~\ref{Masses_table}. There, we substitute 
the estimation of $\kappa$ with that of the characteristic mass $M_*=\sqrt{4\kappa^2}$, equal to the mass 
of the ground vector state -- the lightest massive state in our spectrum. We take the values of $a$ saturating 
the $S$-bound, thus, these are the minimal estimations for $M_*$. Should it be found that $S$ is $p$ times 
smaller, our evaluations for $M_\ast$ become roughly $p$ times larger. For a given set of $\frac{R}{g_5^2}$ 
and $\sin\theta$ lower values of $a$ are permitted and result in larger $M_*$. In addition,  larger $a$ 
leads to  larger splitting between vector fields aligned in different (unbroken and broken) directions. 
It is evident from Table~\ref{Masses_table} that the splitting almost disappears starting from 
$\frac{R}{g_5^2}=10$ for the demonstrated values of $\sin\theta$. We also notice that the effective 
``$N_{hc}$-infinity'' is heralded by the degenerate vector masses in the unbroken and broken sectors 
and starts rather early because $\frac{R}{g_5^2}=10$ fit brings similar results to, say, $\frac{R}{g_5^2}=1000$. 
It is an interesting observation, because in the original AdS/CFT conjecture the strongly coupled Yang--Mills 
theory on the 4D side of the correspondence should be in the limit $N_c\gg1$. Of course, in phenomenological 
AdS/QCD models the duality is commonly extended for the finite values of $N_c$, so we take into consideration 
a set of smaller $\frac{R}{g_5^2}$ as well.

\begin{figure}[t]
	\centering
		\includegraphics[width=0.95\textwidth]{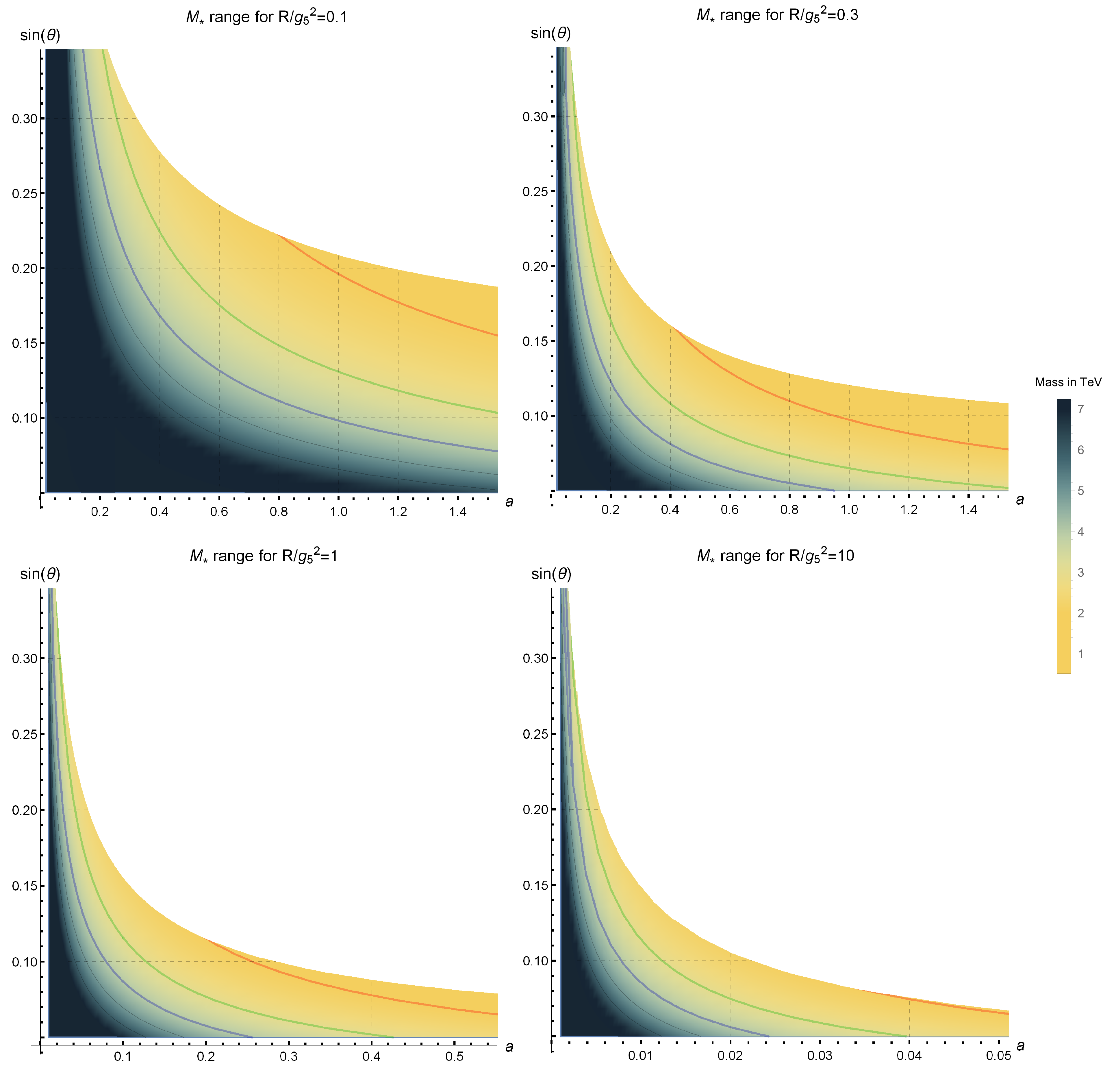}
	\caption{\label{Density_plots}  The density plots of $M_*$ for different values of $R/g_5^2$.
		The colored curves represent the lines of constant $M_*$: the red one -- $M_*=2$~TeV, 
		the green one -- $M_*=3$~TeV, the blue one -- $M_*=4$~TeV and successive black curves for higher 
                integer values. The white area 
		represents the sector prohibited by the $S$ bound.
	}
\end{figure}

In Fig.~\ref{Density_plots} we depict a broader range of $M_*$ values. The dependencies on the model 
parameters could be easily traced from there. In the parameter space $(\sin\theta, a, \frac{R}{g_5^2})$ 
we can fix any two values, then the growth of the third parameter results in lower $M_*$ (as long as it 
does not appear in the prohibited zone). Pursuing higher degree of breaking $a$ results in unlikely small 
masses in the areas that are not well-restrained by the $S$ parameter. We speak of  masses below $2$~TeV 
at smaller values of $\frac{R}{g_5^2}$ and $\sin\theta$. Higher values of other two parameters are more 
efficiently cut off by the $S$ bound. In general, $2.0-4.0$~TeV states are expected. 
We also recollect that in a tower of resonances of one type we have a square root growth with the 
number of a resonance. Thus, for a lowish value of $M_*$ there is a tower with several comparatively
low-lying states. For instance, for the input set $(\sin\theta, a, R/g_5^2)=(0.1, 2.2, 0.3)$ we 
have $M_*=1.3$~TeV and the tower masses are $M_V(n)=\{1.3, 1.8, 2.3, 2.6, \ldots\}$~TeV.

\begin{figure}[t]
	\centering
	\includegraphics[width=0.65\textwidth]{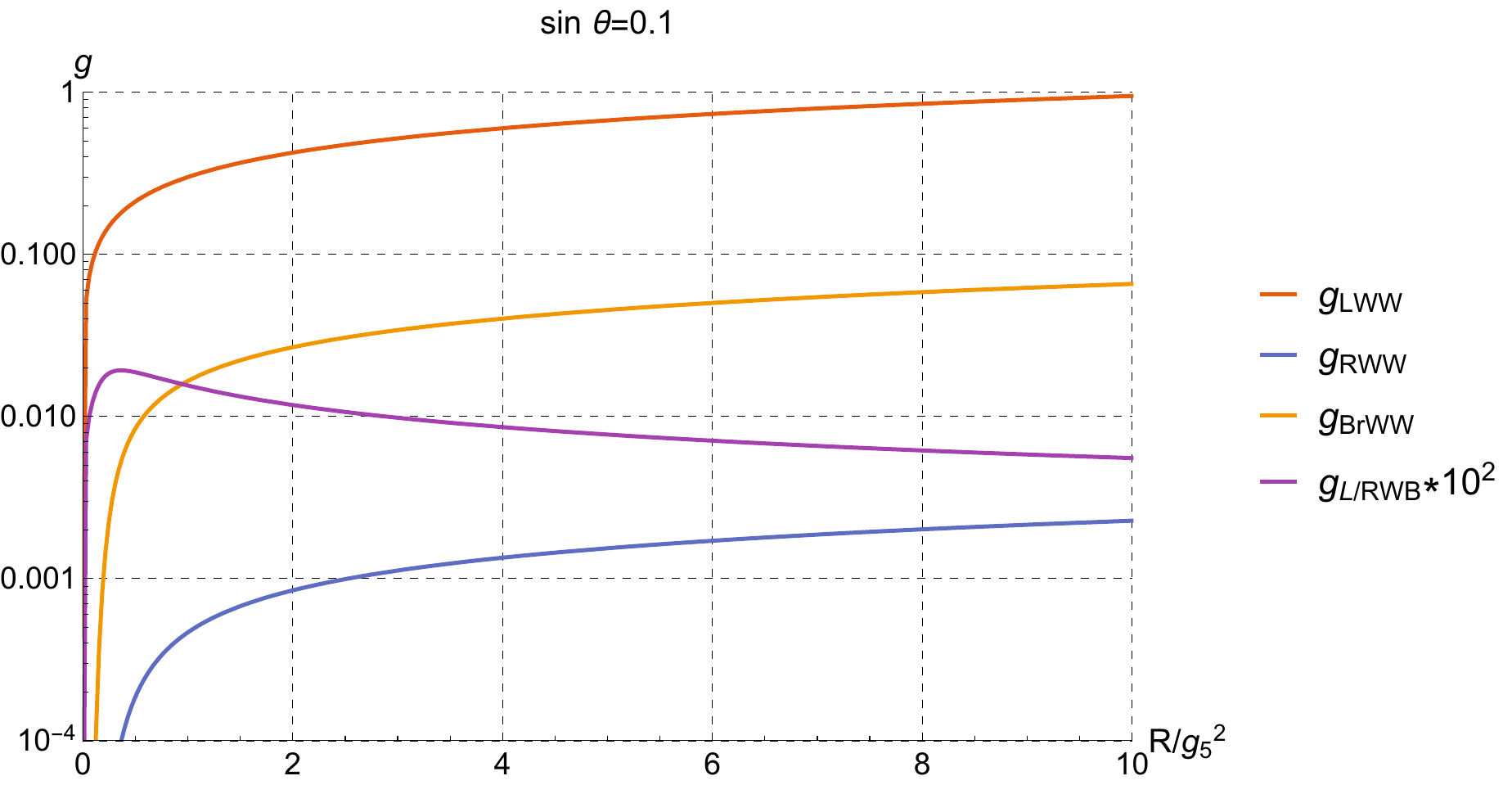}

		\includegraphics[width=0.65\textwidth]{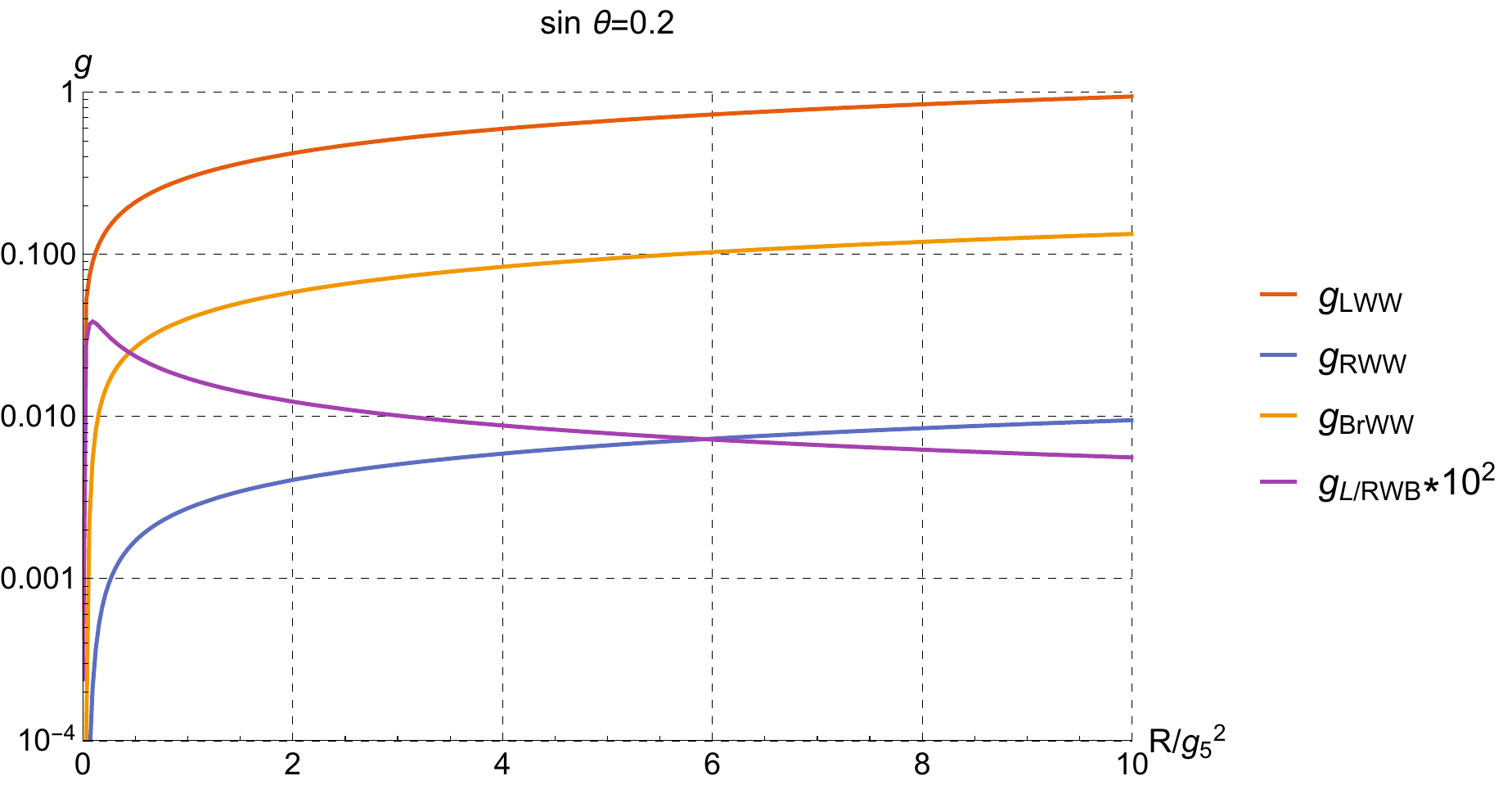}
		
		\includegraphics[width=0.65\textwidth]{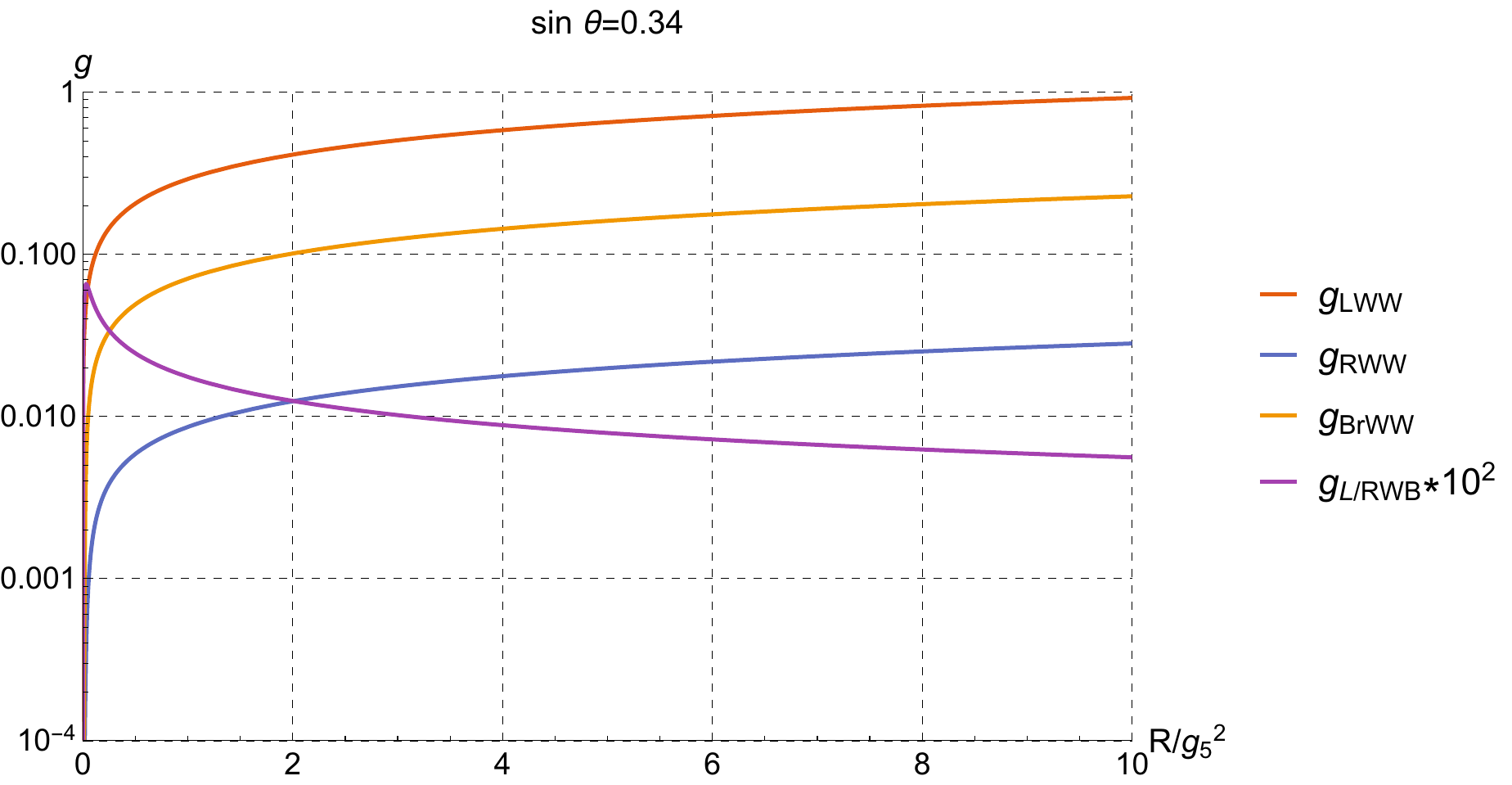}
	
	\caption{\label{Coupl_plots}  Couplings of the left, right and broken composite resonances to the $W^+W^-$ and $W^\pm B$ pairs.
	}
\end{figure}

In Fig.~\ref{Coupl_plots} we present the numerical analysis resulting from Eqns.~(\ref{g_unbr_simpl}), 
(\ref{g_br_simpl}) and (\ref{g_WB_simpl}), showing the possible values of the couplings between the 
left, right and broken resonances and a $W^+W^-$ or $W^3B$-pair.  It is clear that the left 
resonances couple more strongly than  the right ones thanks to the dampening the latter get 
with $\cos\theta$ being rather close to $1$.  All the $WW$ couplings exhibit a logarithmic growth 
with $\frac{R}{g_5^2}$. 
 The parameter $a$ was taken to be saturating the $S$-bound of Fig.~\ref{Splot} and 
is rendered quite close to zero at higher values of $R/g_5^2$ especially for larger $\sin\theta$. 
The coupling including the $B$ meson is rather small in comparison to the $WW$ ones due to the direct 
proportionality to $a$, and it vanishes exactly for $a=0$.
 
\begin{figure}
	\centering
\includegraphics[width=0.65\textwidth]{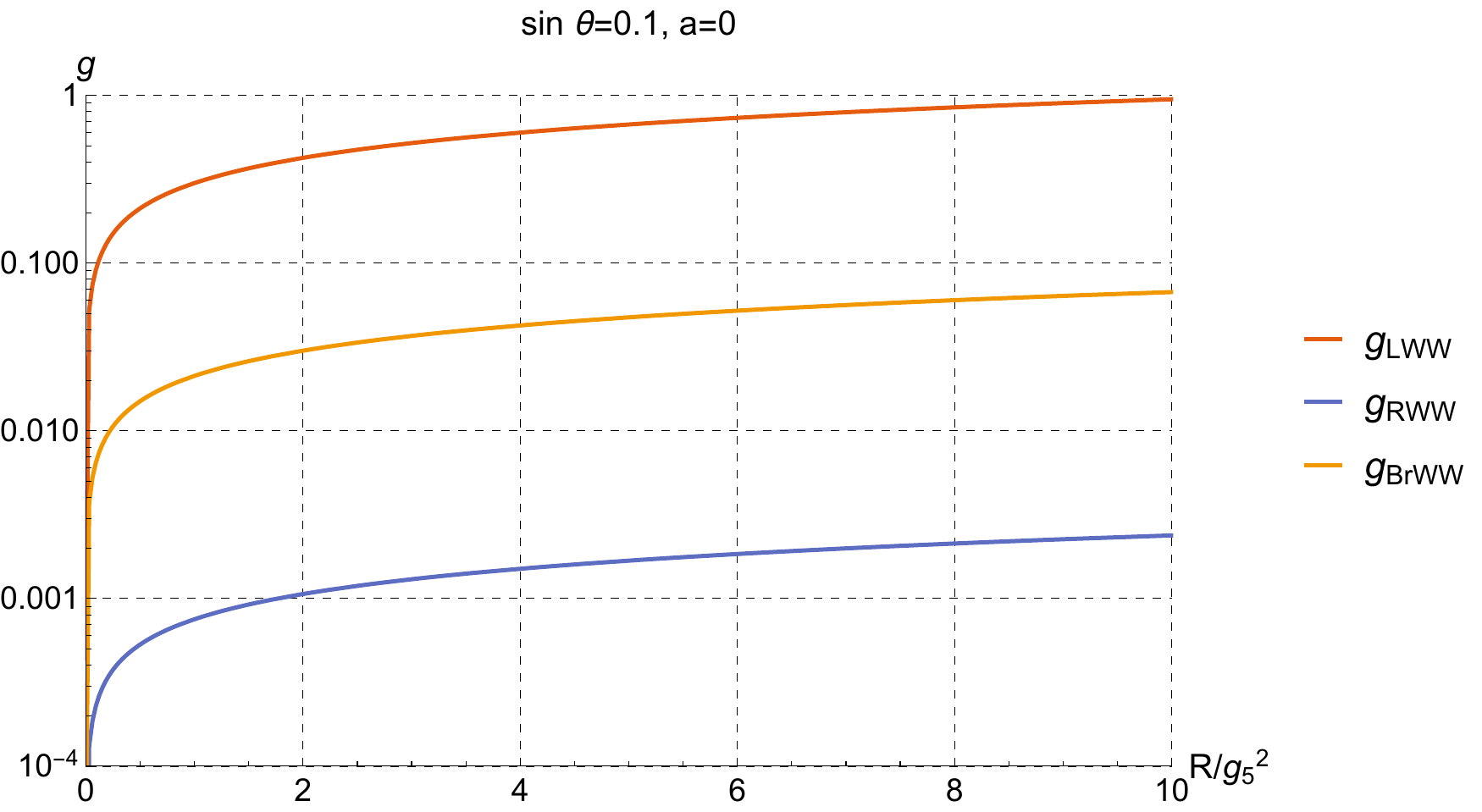}
	\caption{\label{Coupl_zeroa_plot}  Example of the couplings estimated for a completely vanishing value of $a$.
	}
\end{figure}

In order to show the impact of $a$ on $WW$ couplings in more detail we provide the same computation 
in Fig.~\ref{Coupl_zeroa_plot} imposing $a=0$ by hand for the fit with  $\sin\theta=0.1$ 
(the most illustrative case). The difference between this and the top panel of Fig.~\ref{Coupl_plots} is only 
noticeable for $R/g_5^2\lesssim0.5$; and now the saturation is reached sooner.  
At the major part of the $R/g_5^2$ axis the scale of $SO(5)$ breaking is of little consequence 
for the couplings discussed. 
The importance of the $S$ constraint at very small values of $R/g_5^2$ is doubtful. At the same time, 
this area turns out relevant if we assume that the CH value is close to the QCD one, or if we take 
into account the estimations of these couplings made in other studies.

It is not easy to make comparison between the values of the couplings obtained here
and possible experimental bounds because in the analyses of the LHC experimental data on resonances 
decaying into $WW$ or $WZ$ pairs some benchmark signal models are normally used 
(Kaluza--Klein graviton in extra dimension, extended gauge model of $W'$ and $Z'$, and others). 
However, in a more model-independent framework of Ref.~\cite{Delgado2017} we find that the characteristic 
scale for the couplings is of order $0.001 \div 0.010$. $g_{LWW}$ and $g_{BrWW}$ tend to be much larger 
unless computed at very small $R/g_5^2$. We can only speculate about the effect of including quantum 
corrections in our calculation. Barring large corrections, the comparison with Ref.~\cite{Delgado2017} really
indicates lowish values for $R/g_5^2$.

\section{Conclusions}
In this study we used the bottom-up holographic approach to have a fresh look at non-perturbative aspects of CH models with a global 
breaking pattern $SO(5)\to SO(4)$ and a gauge group misaligned with the unbroken group. With the purpose of being as close as possible to
the characteristics of a confining theory (presumed to be underlying the EWSBS) we chose to work
in a 5D SW framework inspired by effective models of QCD and consisting in a generalized 
sigma model coupled both to the composite resonances and to the SM gauge bosons. The 5D model is similar 
to that of successful AdS/QCD constructions, specifically to our earlier work~\cite{Espriu_2020}, and 
depends on the two ansatze functions: the SW dilaton profile $\Phi(z)$ and the symmetry-breaking $f(z)$.
The microscopic nature of the breaking, besides being triggered by some new strong interactions with an 
hyper-color group, is factored out and every effort have been taken to make predictions as independent of
it as possible.

 We investigated the dynamics of ten vector (unbroken and broken) and four Goldstone (one of them related 
to the Higgs) 5D fields. Though for the unbroken vectors the situation is much similar to a generic AdS/QCD model, 
in the broken sector we have developed a procedure that relates the Goldstone 
fields to the fifth component $A_z^i$. That is not just a gauge-Higgs construction because there are as 
well definite independent Goldstone modes in the bulk. The resulting Goldstone description is quite different 
from that of the vector fields. The proposed procedure is  ratified by the agreement of the $hWW$ and $hhWW$ 
characteristic couplings to those of the general MCHM. The Higgs remains massless  
as long as we do not take into account the quantum corrections.

In the paper we lay emphasis on the following issues of phenomenological interest:
\begin{itemize}
	\item  derivation of the spectra of the new states in the broken and unbroken channels;
	\item connection to the EW sector (masses of the gauge bosons and electroweak precision 
observables);
	\item triple couplings of the new heavy resonances to $W^+W^-$ and $W^\pm B$;
	\item in-depth analysis of the realization of the first and second Weinberg sum rules and the study of their convergence.
\end{itemize}

The holographic effective theory describes the composite resonances; their maximum number $N_{max}$ is found 
to be related to the theory natural UV cut-off $\varepsilon$. Adhering to one of these cut-offs is necessary 
to derive relations involving resonance decay constants and masses. The latter stay cut-off independent as 
befits physical observables. The only but very significant exception is the ``pion decay constant'' $F$. 
We made a hypothesis that $\varepsilon$ can be taken as related to the characteristic range of the CH effective 
theory, and provided numerical estimations for the value of $N_{max}$. Moreover, the two Weinberg sum rules 
hold their validity just in a formal sense as the sum over resonances has to be cut off. The sum rules are 
logarithmically divergent, and this implies that they are not saturated at all by just the first resonance. 
We believe it to be a common feature of AdS/CFT models, detached from the particularities of our setup, as
it is also present in holographic QCD. 
We can regard it as a general serious flaw of the bottom-up holographic models, and hence a realistic CH 
theory could also have the sum rules more similar to those of actual QCD.

The minimal set of input parameters in our model is: $\sin\theta$, $a$, and $\frac{g_5^2}{R}$. There are 
constraints coming from the $W$ mass (EW scale), the $S$ parameter and the existing experimental bounds on 
$\kappa_V$ ($\sin\theta$). Their consideration allows us to estimate the masses for the composite resonances. 
It is not difficult to find areas in the parameter space where a resonance between $2$ and $3$~TeV is easily accommodated. 
The presented technique offers the possibility of deriving trilinear couplings of a type $WW$, $WB$--new composite resonance. 
They are of interest because the SM gauge boson scattering is regarded as the process for the new vector resonance
production in collider experiments.

It is compelling to extend the proposed framework to other non-minimal symmetry breaking patterns, especially 
the ones that could be supported by a non-exotic theory at the microscopic level. Then, it would be reasonable 
to include more quantities of physical interest into the analysis.

\begin{acknowledgments}
We acknowledge financial support from the following grants: FPA2016-76005-C2-1-P and  PID2019-105614GB-C21 (MICINN),
and 2017SGR0929 (Generalitat de Catalunya). A.K. acknowledges the financial support of the fellowship BES-2015-072477.
The activities of ICCUB are supported by a Maria de Maeztu grant.

\end{acknowledgments}

\bibliography{biblio/composite_higgs_database}

\appendix
\section{Confluent hypergeometric equation and its solutions}\label{hyperf}
The confluent hypergeometric equation is given as
\be
y \varphi''(y)+(c-y)\varphi'(y)-a\varphi(y)=0.\ee
The values of  parameters $a$ and $c$ define the types of solution one would get~\cite{bateman1}. We abstain from 
considering solutions whose IR asymptotics tend to explode.

For the positive integer values $c=1, 2, 3, ...$ we have 
\be \varphi(y)=C_1\ _1F_1(a,c;y)+C_2 \Psi(a,c;y),\ee where $\ _1F_1(a,c;y)$ is called the Kummer function
and $\Psi(a,c;y)$ is the Tricomi function.

However, in the paper we frequently meet the cases of non-positive  integer $c$. $_1F_1(a,c;y)$ has poles at  
$c=0, -1, -2, ...$, while the Tricomi function can generally be analytically continued to any integer value of $c$.
In that situation we can choose another two solutions from the fundamental system of solutions:
\be\label{hyperfsol} \varphi(y)=C_1 y^{1-c}\ _1F_1(a-c+1,2-c;y)+C_2 \Psi(a,c;y).\ee

Let us discuss several properties of these confluent hypergeometric functions~\cite{bateman1}:
\begin{itemize}
	\item The Tricomi functions with different arguments are related via
	\be \label{Tricrel} \Psi(a,c;y)=y^{1-c}\Psi(a-c+1,2-c;y).\ee
	\item The Tricomi function exhibits a logarithmic behavior for all integer values of $c$.
	Specifically, for the case $c=1-n,\ n=0,1,2,...$ one has
	\begin{align}\notag
	\Psi(a,1-n;y)&=\frac{(n-1)!}{\Gamma(a+n)}\sum\limits_{r=0}^{n-1}\frac{(a)_ry^r}{(1-n)_r r!}
	+\frac{(-1)^{n-1}}{n!\Gamma(a)}\bigg( \ _1F_1(a+n,n+1;x)y^n \ln y+\\ \label{Triclog}
	&+\left.\sum\limits_{r=0}^\infty  \frac{(a+n)_r}{(n+1)_r}[\psi(a+n+r)-\psi(1+r)-\psi(1+n+r)]\frac{y^{n+r}}{r!}\right),
	\end{align}
	here the Pochhammer symbol is $(a)_n=1\cdot a\cdot(a+1)...(a+n-1)=\Gamma(a+n)/\Gamma(a)$,
	$\psi(a)$ is the digamma function; and the first sum is absent for the case $n=0$.
	\item The Tricomi function has an infinite sum representation involving the generalized Laguerre polynomials
	\be\label{app:inf_series_Tricomi}
	\Gamma(a)\Psi(a,1+m;y)=\sum\limits_{n=0}^\infty \frac{L^m_n(y)}{n+a}.\ee
	\item
	The Kummer function is a (finite) series solution 
	$\ _1F_1(a,c;y)=\sum\limits_{n=0}^\infty \frac{(a)_n}{(c)_n}\frac{y^n}{n!}$, that
	has a natural connection with the generalized Laguerre polynomials (for integer $n>0,\ m>0$)
	\be L_n^m(y)=\frac{(m+1)_n}{n!}\ _1F_1(-n,m+1,y).\ee
\end{itemize}

\section{Derivation of the EOM in the broken sector with $\xi=1$}\label{xi1EOM}
Let us assume $A_z^i= \frac{\partial_z \pi^i}{\chi_\pi}$ directly in Eqns.~(\ref{EOM1})-(\ref{EOM3}). 
Then, the system on $A^{i}_\mu$ and $\pi^i$ simplifies to
\begin{align}
\partial_z \frac{e^{-\Phi(z)}}{z}\partial_z A^{i}_\mu&-\frac{e^{-\Phi(z)}}{z}\Box A^{i}_\mu
-\frac{2g_5^2 f^2(z)R^2}{k_s}\frac{e^{-\Phi(z)}}{z^3} A^i_\mu \nn\\ 
&-\partial_\mu\left(\partial_z \frac{e^{-\Phi(z)}}{z}\partial_z \frac{\pi^i}{\chi_\pi}
-\frac{2g_5^2f^2(z)R^2}{k_s}\frac{e^{-\Phi(z)}}{z^3}\frac{ \pi^i}{\chi_\pi}\right)=0 \label{EOMapp1}\\
&\partial^\mu A^{i}_\mu=\Box \frac{\pi^i}{\chi_\pi} \label{EOMapp2}
\end{align}
The condition of Eqn.~(\ref{trans_cond}) holds, and together with Eqn.~(\ref{EOMapp2}) it implies that
\be
\Box^2 \frac{\pi^i}{\chi_\pi}=0.
\ee

With the use of the identity $A^{i\parallel}_\mu=\frac{\partial_\mu\partial^\nu}{\Box}A^{i}_\nu=\partial_\mu \frac{\pi^i}{\chi_\pi}$, 
the longitudinal part in Eqn.~(\ref{EOMapp1}) transforms into
\begin{align}\nn
\partial_z \frac{e^{-\Phi(z)}}{z}\partial_z A^{i\parallel}_\mu&-\frac{2g_5^2 f^2(z)R^2}{k_s}\frac{e^{-\Phi(z)}}{z^3} A^{i\parallel}_\mu \\
&-\partial_\mu\left(\partial_z \frac{e^{-\Phi(z)}}{z}\partial_z \frac{\pi^i}{\chi_\pi}+\frac{e^{-\Phi(z)}}{z}\Box \frac{\pi^i}{\chi_\pi}-\frac{2g_5^2f^2(z)R^2}{k_s}\frac{e^{-\Phi(z)}}{z^3}\frac{ \pi^i}{\chi_\pi}\right)=0.
\end{align}
All things considered, one of the possible solutions is this set of simultaneously fulfilled equations
\begin{align}
&\partial_z \frac{e^{-\Phi(z)}}{z}\partial_z A^{i\parallel}_\mu-\frac{2g_5^2 f^2(z)R^2}{k_s}\frac{e^{-\Phi(z)}}{z^3} A^{i\parallel}_\mu =0,\\
&\partial_z \frac{e^{-\Phi(z)}}{z}\partial_z \frac{\pi^i}{\chi_\pi}-\frac{2g_5^2f^2(z)R^2}{k_s}\frac{e^{-\Phi(z)}}{z^3}\frac{ \pi^i}{\chi_\pi}=0,\\
&\Box \frac{\pi^i}{\chi_\pi}=0,
\end{align}
 while the transverse mode keeps being described by Eqn.~(\ref{AbotEOM}).
 
  With this exercise we intend to be reassured that the masslessness of the Goldstones agrees with 
EOMs~(\ref{pionEOM}), (\ref{AbotEOM}) and  (\ref{AparalEOM}) given in the main body of the paper. 

\section{Large $Q^2$ expansion of the correlator $\Pi_{LR}$}\label{Q2exp}

Here we perform the large $Q^2$ expansion of $\Pi_{LR}$ given by
 \begin{align}\label{LR0}
g^2_V \Pi_{LR}(Q^2)=&\frac{R}{2g_5^2}Q^2\sin^2\theta\bigg\{\psi\left(1+\frac{Q^2}{4\kappa^2}\right)
-\psi\left(1+\frac{Q^2}{4\kappa^2}+a\right)\\\nn
&-\frac{4\kappa^2}{Q^2} a\left[\ln\kappa^2\varepsilon^2+
2\gamma_E+\psi\left(1+\frac{Q^2}{4\kappa^2}+a\right)\right]\bigg\},
\end{align}
by means of using the infinite series representation of the digamma function. From the series 
representation of the $\Gamma$-function it could be derived~\cite{bateman1} 
that
\be \label{app:digamma_series}
\psi(1+z)=-\gamma_E+\sum\limits_{n=1}^\infty \frac z{n(n+z)},
\ee and that is
valid for $z\neq-1,-2,\ldots$. For the particular $\psi$'s of Eqn.~(\ref{LR0}) we have
\begin{align}
&\lim\limits_{Q^2\rightarrow\infty}\psi\left(\frac{Q^2}{4\kappa^2}+1\right)=
-\gamma_E+\sum\limits_{n=0}^\infty\frac1{n+1}\sum\limits_{k=0}^\infty\left(\frac{-M^2_V(n)}{Q^2}\right)^k,\\
\notag&\lim\limits_{Q^2\rightarrow\infty}\psi\left(\frac{Q^2}{4\kappa^2}+1
+\frac{(g_5Rf)^2}{2k_s}\right)=-\gamma_E+\left(1+\frac{2\kappa^2(g_5Rf)^2}{k_sQ^2}\right)\sum\limits_{n=0}^\infty\frac1{n+1}
\sum\limits_{k=0}^\infty\left(\frac{-M^2_A(n)}{Q^2}\right)^k,
\end{align}
where for $k=0$ we have $\lim\limits_{N\rightarrow\infty}\sum\limits_{n=1}^N\frac1n
=\ln N+\gamma_E+\mathcal O(1/N)$.

Substitution of the series expansions yields order by order for $g_V^2\Pi_{LR}(Q^2)/Q^2$
\begin{align}
\left(\frac1{Q^2}\right)^0:&\quad \sin^2\theta\frac{R}{2g_5^2}\left(\sum\limits_{n=0}^\infty\frac1{n+1}-\sum\limits_{n=0}^\infty\frac1{n+1}\right);\\
\label{largeQ2-1}\left(\frac1{Q^2}\right)^1:&\quad 4\kappa^2\sin^2\theta\frac{R}{2g_5^2}\sum\limits_{n=0}^\infty(1-1)-\sin^2\theta\kappa^2a\frac{2R}{g_5^2}\left(\ln\varepsilon^2\kappa^2+\gamma_E+\sum\limits_{n=0}^\infty \frac1{n+1}\right);\\
\label{largeQ2-2}\left(\frac1{Q^2}\right)^2:&\quad 4\kappa^4\sin^2\theta a \frac{2R}{g_5^2}\sum\limits_{n=0}^\infty(1-1).
\end{align}
Considering that $1$ and $-1$, as well as the fractions in the difference between harmonic sums, appear 
together for any fixed $n$ we can set these terms to zeros (certainly $0$ for a finite sum). 
The remaining at $1/Q^2$ order parentheses  cancel due to Eqn.~(\ref{regulator_connection}) when 
the infinite sum is replaced with the one up to $N_{max}$. Thus, we show that the terms $1/Q^2$ 
and $1/Q^4$ are absent as long as $N_{max}<\infty$.

\section{Calculations related to the couplings of Higgs to EW bosons}\label{higgs_coupl_var}
We can factorize the misalignment in Eqns. (\ref{part_hww}) and (\ref{part_hhww}), and come to the following equation
\begin{align}
	\mathcal L_{eff}\supset &\frac{g^2}{g_V^2}\frac{\sin 2\theta}{8\sqrt2}h(q) W^{\alpha}_{\mu}(k_1)W^{\beta}_\nu(k_2) \left[ \frac{\delta^2 S^{(3)}_{5D}}{\delta \phi^\alpha_{L\mu}(k_1)\delta \phi^\beta_{br\nu}(k_2)h(q)} + \frac{\delta^2 S^{(3)}_{5D}}{\delta \phi^\alpha_{br\mu}(k_1)\delta \phi^\beta_{L\nu}(k_2)h(q)} \right] \label{triple_unbr}\\
	&+\frac{g^2}{4g_V^2}h(q_1)h(q_2) W^{\alpha}_{\mu}(k_1)W^{\beta}_\nu(k_2) \left[\cos^2\theta \frac{\delta^2 S^{(4)}_{5D}}{\delta \phi^\alpha_{L\mu}(k_1)\delta \phi^\beta_{L\nu}(k_2)h(q_1)h(q_2)}\right. \label{quartic_unbr}\\ 
	&\left.+ \frac{\sin^2\theta}{2}\frac{\delta^2 S^{(4)}_{5D}}{\delta \phi^\alpha_{br\mu}(k_1)\delta \phi^\beta_{br\nu}(k_2)h(q_1)h(q_2)}\right].\label{quartic_br}
\end{align}
We have made use of the symmetry of the Lagrangian permitting to substitute $\langle h|{\mathcal O}^\alpha_{L\ \mu}{\mathcal O}^\beta_{br\ \nu}|0\rangle=-\langle h|{\mathcal O}^\alpha_{R\ \mu}{\mathcal O}^\beta_{br\ \nu}|0\rangle$ and $\langle hh|{\mathcal O}^\alpha_{L\ \mu}{\mathcal O}^\beta_{L\ \nu}|0\rangle=\langle hh|{\mathcal O}^i_{R\ \mu}{\mathcal O}^\beta_{R\ \nu}|0\rangle=-\langle hh|{\mathcal O}^\alpha_{L\ \mu}{\mathcal O}^\beta_{R\ \nu}|0\rangle$.

Let us explore the triple coupling first. The 5D action provides two types of contributions
\bea \nn
\frac{\delta^2 S^{(3)}_{5D}}{\delta \phi^\alpha_{L\mu}(k_1)\delta \phi^\beta_{br\nu}(k_2)h(q)}
=\delta^{\alpha\beta}\eta_{\mu\nu} \frac{R}{g_5^2}\left(a\kappa^2 \int dy \frac{e^{-y}}{y} \pi(y)/\chi_\pi V(k_1,y) A(k_2,y)\right.\\
\left.+\frac14 \int dy  \frac{e^{-y}}{y}  \partial_z\pi(y)/\chi_\pi V(k_1,y) \partial_zA(k_2,y) 
\right), \label{app:3ptvar}
\eea
and the second variation in (\ref{triple_unbr}) evaluates the same but for exchange $k_1 \leftrightarrow k_2$.

Further, we would like to integrate analytically over $y$. As we substitute the Goldstone profile and the longitudinal vector propagators, all dependence on momenta disappears and the calculation can be performed. For the transverse modes we put the propagators  on-shell with $k_1^2=k_2^2=M^2_W$ and consider the limit $M^2_W\ll 4\kappa^2$. 
 Indeed, we naturally expect the composite resonances to have rather large masses and that limit is substantiated numerically 
in Section~\ref{phen}. Essentially, we set $k_1^2=k_2^2=0$, and the outcoming integral is analogous to the expression with the longitudinal propagators.

 In the calculation it is convenient to use the definitions in terms of the resonance sums
\bea \nn
A(0,z)= F \pi(z) /\chi_\pi = \Gamma(1+a)\Psi(a,0;\kappa^2z^2)=\sum_n \frac{\kappa^2z^2 L^1_n(\kappa^2z^2)}{n+1+a},\\
\partial_z A(0,z)= F \partial_z \pi(z) /\chi_\pi = 2\kappa^2z (-a)\Gamma(1+a)\Psi(a+1,1;\kappa^2z^2)=-2\kappa^2za\sum_n \frac{L_n(\kappa^2z^2)}{n+1+a}. \nn
\eea
Then, the variation (\ref{app:3ptvar}) could be estimated quite easily due to the orthogonality of the Laguerre polynomials
\bea
\kappa^2 a F^{-1} \frac{R}{g_5^2}\sum_{n_1,n_2} 
\frac{\int dy e^{-y} y L^1_{n_1}(y)L^1_{n_2}(y)+a\int dy e^{-y} L_{n_1}(y)L_{n_2}(y)}{(n_1+a+1)(n_2+a+1)}\\=\frac1{2F}\frac{2R\kappa^2 a }{g_5^2}\sum_{n_1,n_2}  \delta_{n_1n_2}\frac{n_1+1+a}{(n_1+a+1)(n_2+a+1)}=\frac F2.
\eea
Here we used for $F^2$ the definition of Eqn.~(\ref{FinSum}).

We follow the same lines for the quartic couplings. 
Let us start with the variation in (\ref{quartic_unbr}):
\begin{align}
\frac{\delta^2 S^{(4)}_{5D}}{\delta \phi^\alpha_{L\mu}(k_1)\delta \phi^\beta_{L\nu}(k_2)h(q_1)h(q_2)}=&2\delta^{\alpha\beta}\eta_{\mu\nu} \frac{R}{4 g_5^2}\left(a\kappa^2 \int dy \frac{e^{-y}}{y} (\pi(y)/\chi_\pi)^2 V(k_1,y) V(k_2,y)\right.\\
&\left.+\frac14 \int dy  \frac{e^{-y}}{y}  (\partial_z\pi(y)/\chi_\pi)^2 V(k_1,y) V(k_2,y)
\right) \nn \\
&= \frac14 \delta^{\alpha\beta}\eta_{\mu\nu} F^{-2} \frac{2R}{g_5^2}a\kappa^2 \sum_n \frac{n+1+a}{(n+1+a)^2}= \frac14 \delta^{\alpha\beta}\eta_{\mu\nu}.
\end{align}
Unfortunately, the situation becomes more involved  with the variation over the broken sources 
in (\ref{quartic_br}) because the integrals there are quartic in Laguerre polynomials 
\begin{align}
\frac{\delta^2 S^{(4)}_{5D}}{\delta \phi^\alpha_{br\mu}(k_1)\delta \phi^\beta_{br\nu}(k_2)h(q_1)h(q_2)}
&=\delta^{\alpha\beta}\eta_{\mu\nu} F^{-2} \frac{R}{g_5^2}a\kappa^2 \nn \\
&\times \sum_{n_1,n_2}\frac{\int dy e^{-y} A^2(0,y) [a/2 L_{n_1}(y)L_{n_2}(y) - y L^1_{n_1}(y)L^1_{n_2}(y)]}{(n_1+a+1)(n_2+a+1)}
\end{align}
We can make a calculation at $a=0$
, with the result  
$\frac{\delta^2 S^{(4)}_{5D}}{\delta \phi^\alpha_{br\mu}(k_1)\delta \phi^\beta_{br\nu}(k_2)h(q_1)h(q_2)}= -\frac12 \delta^{\alpha\beta}\eta_{\mu\nu}$.
We extrapolate this estimation to the case of general $a$ when we present the quartic coupling in the effective Lagrangian.

\section{Calculations related to the couplings of vector resonances to EW bosons}\label{resonance_coupl_var}
Here we calculate the relevant three-point functions first. Diagrammatically, we obtain a vertex and three propagators with 
their residues attached to it. In the body of the paper we report the effective vertex proceeding from connecting two 
legs to the physical sources and reducing the third one via putting an $n$-th resonance on-shell.

There are not that many types of different three-point functions that can be extracted from Eqn.~(\ref{triple_int_A})
\begin{align}
&\langle \mathcal O^{\alpha}_{L\mu_1}(q_1)\mathcal O^{\beta}_{L\mu_2}(q_2)\mathcal O^{\gamma}_{L\mu_3}(q_3)\rangle
=\langle \mathcal O^{\alpha}_{R\mu_1}(q_1)\mathcal O^{\beta}_{R\mu_2}(q_2)\mathcal O^{\gamma}_{R\mu_3}(q_3)\rangle\\\nn
&=i\varepsilon^{\alpha\beta\gamma}\texttt{Lor}_{\mu_1\mu_2\mu_3}\delta(q_1+q_2+q_3)T_{3V}(q_1,q_2,q_3);\\
&\langle \mathcal O^{\alpha}_{L\mu_1}(q_1)\mathcal O^{\beta}_{br\mu_2}(q_2)\mathcal O^{\gamma}_{br\mu_3}(q_3)\rangle
=\langle \mathcal O^{\alpha}_{R\mu_1}(q_1)\mathcal O^{\beta}_{br\mu_2}(q_2)\mathcal O^{\gamma}_{br\mu_3}(q_3)\rangle\\\nn
&=i\varepsilon^{\alpha\beta\gamma}\texttt{Lor}_{\mu_1\mu_2\mu_3}\delta(q_1+q_2+q_3)\frac 12 T_{V2A}(q_1,q_2,q_3);\\
&\langle \mathcal O^{4}_{br\mu_1}(q_1)\mathcal O^{\alpha}_{br\mu_2}(q_2)\mathcal O^{\beta}_{R\mu_3}(q_3)\rangle
=-\langle \mathcal O^{4}_{br\mu_1}(q_1)\mathcal O^{\alpha}_{br\mu_2}(q_2)\mathcal O^{\beta}_{L\mu_3}(q_3)\rangle\\\nn
&=i\delta^{\alpha\beta}\texttt{Lor}_{\mu_1\mu_2\mu_3}\delta(q_1+q_2+q_3)\frac 12 T_{V2A}(q_3,q_1,q_2).
\end{align}
There, the Lorentz structure of the correlators is collected into 
$$ \texttt{Lor}_{\mu_1\mu_2\mu_3}(q_1,q_2,q_3)=\eta_{\mu_1\mu_2}(q_1-q_2)_{\mu_3}+\eta_{\mu_1\mu_3}(q_3-q_1)_{\mu_2}
+\eta_{\mu_2\mu_3}(q_2-q_3)_{\mu_1},$$
and we defined the form factors as follows
\bea
T_{3V}(q_1,q_2,q_3)=\frac R{g_5^2}\int dz e^{-\kappa^2z^2} z^{-1} V(q_1,z)V(q_2,z)V(q_3,z),\\
T_{V2A}(q_1,q_2,q_3)=\frac R{g_5^2}\int dz e^{-\kappa^2z^2} z^{-1} V(q_1,z)A(q_2,z)A(q_3,z).
\eea

Now, to consider the possible interactions with $W$ and $B$ bosons we write down the relevant three-point functions
\begin{align}\nn
\langle \mathcal O^{\alpha}_{L/R\mu_1}(q_1)\widetilde{J}^{\beta}_{L\mu_2}(q_2)\widetilde{J}^{\gamma}_{L\mu_3}(q_3)\rangle =&
\frac{g^2}{8g_V^2} i\varepsilon^{\alpha\beta\gamma}\texttt{Lor}_{\mu_1\mu_2\mu_3}(q_1,q_2,q_3)\delta(q_1+q_2+q_3) 
\\& \times\left[(1\pm\cos\theta)^2 T_{3V}(q_1,q_2,q_3)+ \sin^2\theta T_{V2A}(q_1,q_2,q_3)\right];
\\ \nn
\langle \mathcal O^{\alpha}_{L/R\mu_1}(q_1)\widetilde{J}^{\beta}_{L\mu_2}(q_2)\widetilde{J}^{3}_{R\mu_3}(q_3)\rangle =& 
\langle \mathcal O^{\alpha}_{L/R\mu_1}(q_1)\widetilde{J}^{3}_{R\mu_2}(q_2)\widetilde{J}^{\beta}_{L\mu_3}(q_3)\rangle\\
&=\frac{gg'}{8g_V^2} i\varepsilon^{\alpha\beta 3}\texttt{Lor}_{\mu_1\mu_2\mu_3}(q_1,q_2,q_3)\delta(q_1+q_2+q_3) \nn
\\& \times\left[(1-\cos^2\theta) T_{3V}(q_1,q_2,q_3)- \sin^2\theta T_{V2A}(q_1,q_2,q_3)\right];\\ 
\langle \mathcal O^{\alpha}_{br\mu_1}(q_1)\widetilde{J}^{\beta}_{L\mu_2}(q_2)\widetilde{J}^{\gamma}_{L\mu_3}(q_3)\rangle =&-
\frac{g^2}{2g_V^2}\frac{\sin\theta}{\sqrt 2}i\varepsilon^{\alpha\beta\gamma}\nn
\texttt{Lor}_{\mu_1\mu_2\mu_3}(q_1,q_2,q_3) \delta(q_1+q_2+q_3)\\
&\times\left[T_{V2A}(q_2,q_1,q_3)+T_{V2A}(q_3,q_2,q_1)\right] ;\\
\nn
\langle \mathcal O^{\alpha}_{br\mu_1}(q_1)\widetilde{J}^{\beta}_{L\mu_2}(q_2)\widetilde{J}^{3}_{R\mu_3}(q_3)\rangle 
=&\frac{gg'}{2g_V^2} \frac{\sin\theta}{\sqrt 2} i\varepsilon^{\alpha\beta 3}\texttt{Lor}_{\mu_1\mu_2\mu_3}(q_1,q_2,q_3)\delta(q_1+q_2+q_3) \nn
\\& \times\left[T_{V2A}(q_2,q_1,q_3)-  T_{V2A}(q_3,q_2,q_1)\right];
\\
\langle \mathcal O^{4}_{br\mu_1}(q_1)\widetilde{J}^{\alpha}_{L\mu_2}(q_2)\widetilde{J}^{\beta}_{L\mu_3}(q_3)\rangle =&
\frac{g^2 \sin 2\theta}{8\sqrt 2g_V^2}\delta^{\alpha\beta}
\texttt{Lor}_{\mu_1\mu_2\mu_3}(q_1,q_2,q_3)\delta(q_1+q_2+q_3) \nn \\
&\times\left[T_{V2A}(q_3,q_1,q_2)-T_{V2A}(q_2,q_1,q_3)\right] \label{A4_3pt};\\
\langle \mathcal O^{4}_{br\mu_1}(q_1)\widetilde{J}^{3}_{R\mu_2}(q_2)\widetilde{J}^{3}_{R\mu_3}(q_3)\rangle=&
\frac{g'^2 \sin 2\theta}{8\sqrt 2g_V^2} \texttt{Lor}_{\mu_1\mu_2\mu_3}(q_1,q_2,q_3)\delta(q_1+q_2+q_3) \nn \\
&\times\left[T_{V2A}(q_3,q_1,q_2)-T_{V2A}(q_2,q_1,q_3)\right],\\
\langle \mathcal O^{4}_{br\mu_1}(q_1)\widetilde{J}^{3}_{L\mu_2}(q_2)\widetilde{J}^{3}_{R\mu_3}(q_3)\rangle=&
\frac{gg' \sin 2\theta}{8\sqrt 2g_V^2} \texttt{Lor}_{\mu_1\mu_2\mu_3}(q_1,q_2,q_3)\delta(q_1+q_2+q_3) \nn \\
&\times\left[T_{V2A}(q_3,q_1,q_2)-T_{V2A}(q_2,q_1,q_3)\right].
\end{align}
Only a few $BB$--resonance interactions are possible due to the epsilon-tensor on the right-hand side of the holographic three-point functions.

Further, we reduce the leg corresponding to $q_1$ momentum and consider the limit $q^2_{2,3}\ll 4\kappa^2$ for other 
two momenta. For the $n$-th excitation of the left/right resonances in the unbroken sector that means:
\bea
T_{3V}(q_1,q_2,q_3)\rightarrow&  \sqrt{\frac{R}{2g_5^2(n+1)}}\int dy e^{-y}L_n^1(y)=\sqrt{\frac{R}{2g_5^2(n+1)}},\\
T_{V2A}(q_1,q_2,q_3)\rightarrow&  \sqrt{\frac{R}{2g_5^2(n+1)}}\int dy e^{-y}L_n^1(y)\Gamma^2(1+a)\Psi^2(a,0;y),
\eea
where the latter integral can be calculated for a given $n$. For $n=0$: $1-2a+2a^2\psi_1(1+a)$.

For the $n$-th excitation of the resonances from the broken sector one of the broken legs should be reduced, and we get
\bea
T_{V2A}(q_2,q_1,q_3)\ \text{or }T_{V2A}(q_3,q_2,q_1)&\rightarrow&  \sqrt{\frac{R}{2g_5^2(n+1)}}\sum_{n'} \frac{\int dy e^{-y}L_n^1(y)L_{n'}^1(y)}{n'+1+a}\\
&&=\sqrt{\frac{R(n+1)}{2g_5^2}}\frac1{n+1+a}. \nn
\eea

Some triple couplings will not be included in the effective Lagrangian. These are: 
$A^4_{br} W^\alpha W^\alpha$, $A^4_{br} B B$, $A^4_{br} W^3 B$, $A^\alpha_{br} W^\beta B$.  The reason for it is that  
in the corresponding three-point functions the leading term in the limit $q^2_{2,3}\ll 4\kappa^2$ is zero due 
to the subtraction of the form factors. The first contribution is $\sim \frac{M^2_W}{4\kappa^2}$ and, thus, is 
strongly suppressed. We abstain from considering observables of this order in this work.

\end{document}